\documentclass{article}%
\usepackage[margin=1in]{geometry}
\usepackage{amsmath}
\usepackage{amsfonts}
\usepackage{amssymb}
\usepackage{graphicx}%
\begin{document}

\begin{center}
{\Large Fat tails arise endogenously from supply/demand, with or without jump
processes}
\end{center}

{\Large \bigskip}

\begin{center}
{\large Gunduz Caginalp}

University of Pittsburgh

Pittsburgh, PA 15260

caginalp@pitt.edu
\end{center}

\bigskip

\begin{center}
\bigskip
\end{center}

\textbf{Abstract.} We show that the quotient of Levy processes of
jump-diffusion type has a fat-tailed distribution. An application is to price
theory in economics, with the result that fat tails arise endogenously from
modeling of price change based on an excess demand analysis resulting in a
quotient of arbitrarily correlated demand and supply whether or not jump
discontinuities are present. The assumption is that supply and demand are
described by drift terms, Brownian (i.e., Gaussian) and compound Poisson jump
processes. If $P^{-1}dP/dt$ (the relative price change in an interval $dt$) is
given by a suitable function of relative excess demand, $\left(
\mathcal{D}-\mathcal{S}\right)  /\mathcal{S}$ (where $\mathcal{D}$ and
$\mathcal{S}$ are demand and supply), then the distribution has tail behavior
$F\left(  x\right)  \sim x^{-\zeta}$ for a power $\zeta$ that depends on the
function $G$ in $P^{-1}dP/dt=G\left(  \mathcal{D}/\mathcal{S}\right)  $. For
$G\left(  x\right)  \sim\left\vert x\right\vert ^{1/q}$ one has $\zeta=q.$ The
empirical data for assets typically yields a value, $\zeta\tilde{=}3,$ or
$\zeta\in\left[  3,5\right]  $ for many financial markets.

Many theoretical explanations have been offered for the disparity between the
tail behavior of the standard asset price equation and empirical data. This
issue never arises if one models price dynamics using basic economics
methodology, i.e., generalized Walrasian adjustment, rather than the usual
starting point for classical finance which assumes a normal distribution of
price changes. The function $G$ is deterministic, and can be calibrated with a
smaller data set. The results establish a simple link between the decay
exponent of the density function and the price adjustment function, a feature
that can improve methodology for risk assessment.

The mathematical results can be applied to other problems involving the
relative difference or quotient of Levy processes of jump-diffusion type.

\bigskip


\bigskip

\textbf{1. Introduction.} \ The calculation or approximation of a quotient of
random variables or stochastic processes is considerably more important in
mathematical modeling than it first appears. In many physical, biological and
economics systems, there is a competition between two factors, e.g., supply
and demand.\ The difference results in a temporal change in another important
variable such as price. However, it is usually the relative difference that is
relevant, and requires the division by one of the variables. Furthermore, if
$X$ and $Y$ are the two random variables with some dimensions, and $Z$ is a
dimensionless variable, and $\tau$ a time scale, then the equation $\tau
dZ/dt=X-Y$ would not be meaningful from the dimensional analysis perspective.
The right hand side must be divided by either $X,$ $Y$ or a universal
constant, or another variable with the same units. If $X$ and $Y$ are
velocities (and relativistic), for example, the speed of light, $c,$ would be
a possible divisor. However, if there are no natural, universal constants,
then it is likely from a dimensional analysis perspective that division by one
of the variables is needed. This leads to consideration of an equation such as
$\tau dZ/dt=\left(  X-Y\right)  /Y=X/Y-1.$

If $X$ and $Y$ are random variables, then the density of $\tau dZ/dt$ is given
by the shifted density of the pure quotient $X/Y.$ In particular, the density
of a quotient has been studied in the literature for Gaussian random variables
$X$ and $Y$ dating back to 1930 \cite{GE} (see references in \cite{PKM}). The
quotient has also been studied for a number of other distributions \cite{SN}.
In most works, the random variable are assumed to be independent. The results
are interesting in view of the broad range of approximate densities one
obtains, including behavior that is similar to a Gaussian in the mid-range.
Motivated by price changes in economics, Caginalp and Caginalp \cite{CC}
studied the tail density problem for $X$ and $Y$ that are arbitrarily
correlated Gaussian variables, finding an exact expression for the full
density in the special case for which $X$ and $Y$ are anti-correlated. A
closely related formula is proven in the Appendix A of this paper. In this
paper we extend the general quotient results to include processes with jumps.
Among the subtleties is the nature of correlations between the normal terms,
the timing of the jumps and the magnitude of the jumps in the numerator and denominator.

For stochastic processes, $X_{t}$ and $Y_{t}$, one has similar questions. The
mathematical problem we study in this paper is to determine the behavior of
the tail of the distribution of $Z_{t}$ where (incorporating $\tau$ into the
variable $Z),$%
\[
Z_{t}=\frac{X_{t}}{Y_{t}},
\]
and $X_{t}$ and $Y_{t}$ are stochastic processes that consist of Brownian
motion plus jump processes (more precisely, Levy processes of jump-diffusion
type defined below).

The randomness in both $X_{t}$ and $Y_{t}$ is assumed to arise from the
normals, the timing of the jumps, and the magnitude of the jumps.

Throughout this paper we focus on a classic price equation in economics,
thereby addressing the problem through that perspective. However, the results
are easily interpreted for other applications in physics, biology, etc.

\textbf{ }A standard assumption in classical finance has been that relative
changes in asset prices have a normal (Gaussian) distribution. In other words,
with price $P\left(  t\right)  $ as a function of time, $t,$ the price change
$\Delta P\left(  t\right)  :=P\left(  t+\Delta t\right)  -P\left(  t\right)  $
in a time interval, $\Delta t$, satisfies $P^{-1}\Delta P\sim\mathcal{N}%
\left(  \mu,\sigma\Delta t\right)  ,$ i.e., a normal distribution with mean
$\mu$ (the expected return) and variance $\sigma\Delta t.$ This idea dates
back to Bachelier's thesis \cite{BA} in 1900, and was broadly adopted by the
finance community in mid-20th century for the pricing of options (e.g.,
Black-Scholes) and the assessments of risk. A popular idea for risk assessment
has been the use of "value-at-risk" analysis \cite{HW} whereby one seeks to
determine whether an investment will retain at least, say, $66\%$ of its value
with a probability of $90\%,$ for example, during the next decade. In
addressing this question, classical methods use the historical data for that
investment to determine the variance and the assumption of a normal
distribution to calculate the value-at-risk probability.

Black-Scholes options pricing is also based on the idea that relative price
change is normally distributed. In continuum form the basic stochastic
equation is generally written as (see e.g., \cite{BA, BS, HW})
\begin{equation}
P^{-1}dP=\mu dt+\sigma dW, \label{cs}%
\end{equation}
where $W$ is Brownian motion, so that $\Delta W:=W\left(  t+\Delta t\right)
-W\left(  t\right)  \sim\mathcal{N}\left(  0,\Delta t\right)  $ and $\mu$ and
$\sigma$ are the mean and standard deviation of the stochastic process. The
theoretical justification for equation $\left(  \ref{cs}\right)  $ is limited,
and its widespread use is partly attributable to mathematical convenience
\cite{CH, CC} and the fact that it is a basic equation from which one can
build additional models. In fact, one justification is based on infinite
divisibility\footnote{A probability distribution $F$ is infinitely divisible
if for any integer $n\geq2$ there exist independent, identically distributed
random variables $X_{1},..,X_{n}$ such that $X_{1}+..+X_{n}$ has distribution
$F.$ In particular, any Levy process, $\left\{  X_{t}\right\}  $ for any
$t\geq0$ has an infinitely divisible distribution (\cite{CT}, p. 69).}.
Roughly speaking, this means that the random price change in a particular time
interval can be writen as a sum random variables on arbitrarily smaller time
intervals. However, there are many other distributions besides Gaussian that
are also infinitely divisible. So the property of infinite divisibility does
not imply that one can apply the Central Limit Theorem to price changes.

A direct implication of the basic equation $\left(  \ref{cs}\right)  $ is that
the density of the relative price change on a fixed interval, $\Delta t,$ is
normal, e.g., for $\mu:=const,$ one has $P^{-1}\Delta P\sim\mathcal{N}\left(
\mu,\sigma^{2}\Delta t\right)  $ so that the density of the relative price
change is
\[
f\left(  x\right)  =\left(  2\pi\sigma^{2}\Delta t\right)  ^{-1/2}\exp\left\{
-\left(  x-\mu\right)  ^{2}/\left(  2\sigma^{2}\Delta t\right)  \right\}  ,
\]
which exhibits the classical exponential decay for large $x.$

While there is some empirical justification for normality of relative price
changes (often called return in the finance literature), the deviations
between the data and a normal distribution become especially pronounced at the
tails. This phenomenon of the tail of a density decaying at a rate that is a
power law, i.e., much slower than the exponential of the normal distribution,
is often called \emph{fat tails}. A large discrepancy\ has been observed
between the implications for the frequency of unusual events
\cite{R,L,MS,KE,G, GGP, DJ, MA,TD,BP,XA} in the data relative to that obtained
from the normal distribution with parameters calculated from the standard
deviation of a sample. Fat tails have also been observed in Bitcoin
\cite{EKK1}, Korea \cite{EKK2}\ and even in laboratory experiments \cite{KH}
where uninformed traders are mainly responsible for these tails. Studies that
have provided a theoretical framework for fat tails include \cite{M},
\cite{MS} \cite{JV}, \cite{MH}.\ In particular, if one measures the standard
deviation, $\sigma,$ and the drift, or expected return, $\mu$, for the S\&P
500 and utilizes that value in the normal density above, then $\left(
\ref{cs}\right)  $ would imply that a 4\% drop, for example, occurs with a
frequence that is one in many millions of days, instead of one in about 500
days, the observed frequency. Empirical data suggests that for most stock
classifications, one has power law decay, i.e., $f\left(  x\right)  \sim
x^{-\alpha}$ for some $\alpha>0$ (see Conclusion for comparison of our results
with empirical data). The data for goods (see e.g.,\cite{KM}) is less
conclusive. The main objective of this paper is to prove that one obtains a
power law decay from a price model that is based on economic principles of
supply and demand (with any correlation) under conditions that include
stochastic jumps in supply and demand.

The classical approach leading to $\left(  \ref{cs}\right)  $ marginalizes the
issues involving supply and demand, modeling instead the price change $\left(
\ref{cs}\right)  $ as an empirically observed phenomenon. As an alternative to
this semi-empirical equation $\left(  \ref{cs}\right)  ,$ one can model price
dynamics of assets directly through supply and demand, which ultimately drive
price changes. As noted in \cite{CC} a simple and standard model for pricing
of goods is the excess demand model that essentially describes price change in
time as a mechanism to restore equilibrium (e.g. \cite{WG} or \cite{WE}). This
is known in classical economics as Walrasian price adjustment, and is
generally expressed as
\begin{equation}
p_{t}-p_{t-1}=\frac{1}{\tau}\left(  d_{t-1}-s_{t-1}\right)
,\label{discrete p}%
\end{equation}
where $p_{t}$ is the price at the discrete time, $t$, and $s_{t-1}$ and
$d_{t-1}$ are the supply and demand at time $t-1$, and $\tau^{-1}$ is a
constant that determines the extent to which prices move for each unit of
imbalance between supply and demand . The supply and demand functions are
assumed to be linear, which are good approximations for smooth functions when
deviations from the equilibrium point are small.

Of course, $\left(  \ref{discrete p}\right)  $ is only a local equation that
is valid for a particular pair of linear supply and demand near a point of
intersection. For example, an imbalance created by $d_{t-1}=1,010$ and
$s_{t-1}=1,000$ will have a much smaller impact on price change than would
$d_{t-1}=20$ and $s_{t-1}=10$ even though $d_{t-1}-s_{t-1}=10$ in both cases.
This demonstrates the need for normalization, realized by dividing the right
hand side of $\left(  \ref{discrete p}\right)  $ by $s_{t-1}.$ Similarly, the
left hand side must be normalized by dividing by $p_{t-1}$ leading to the
equation, with $\tau_{0}$ now a dimensionless constant that characterizes the
extent to which relative price changes as relative excess demand changes:%
\begin{equation}
\tau_{0}\frac{p_{t}-p_{t-1}}{p_{t-1}}=\frac{d_{t-1}-s_{t-1}}{s_{t-1}}.
\label{lin discrete}%
\end{equation}

The fact that we need to consider proportions of suply and demand implies that
we have, instead of the difference $d_{t-1}-s_{t-1},$ the quotient minus $1,$
i.e., $d_{t-1}/s_{t-1}-1.$ When we consider probability densities, the $-1$
simply produces a shift, so the probabilistic issues reduce to determining the
density of a quotient of random variables.

Note that while these normalizations lead to an equation that is a reasonable
non-local model, another feature of $\left(  \ref{discrete p}\right)  $ is
that it is a linear equation, so that the price change is always proportional
to the excess demand. While linearity is often a convenient and reasonable
approximation, there is no compelling requirement that price change be a
linear function of excess demand. Introducing a differentiable function
$g:\mathbb{R}^{+}\mathbb{\rightarrow R}$ with suitable
properties\footnote{Note that we can write the generalization of the right
hand side of $\left(  \ref{lin discrete}\right)  $ as
\[
\hat{g}\left(  \frac{d_{t-1}-s_{t-1}}{s_{t-1}}\right)  =\hat{g}\left(
\frac{d_{t-1}}{s_{t-1}}-1\right)  =g\left(  \frac{d_{t-1}}{s_{t-1}}\right)  ,
\]
so $g$ is defined as a shift of $\hat{g},$ i.e., $g\left(  x\right)  =\hat
{g}\left(  x-1\right)  .$} including $g\left(  1\right)  =0$ and $g^{\prime
}>0,$ we can write in place of $\left(  \ref{lin discrete}\right)  $ the
equation%
\begin{equation}
\tau_{0}\frac{p_{t}-p_{t-1}}{p_{t-1}}=g\left(  \frac{d_{t-1}}{s_{t-1}}\right)
. \label{nonlin discrete p}%
\end{equation}

Thus, information on the supply and demand at any discrete time determines the
price change for the next discrete time. The design of markets and efficient
price discovery has been an active research area from both a theoretical and
experimental perspective. See models by Milgrom \cite{MI}, Gjerstad and
Dickhaut \cite{GD}, Hirshleifer et. al. \cite{HGH}, Gjerstad \cite{G2},
\cite{G1}, and references therein. The experimental aspect has been studied by
researchers including Plott and Pogorelskiy, \cite{PP}, Bossaerts and Plott,
\cite{BP}, Porter et. al. $\cite{PR}.$

A modern approach to modeling price dynamics as asset flow, initiated in 1990,
has been built on analyzing the supply and demand as discussed above (see e.g.
$\cite{CE},$ \cite{CB} and more recently \cite{MA1}, $\cite{DW}$). In other
words, we build a continuum model based on the same principle as the discrete
models above. The supply and demand can depend on the price, price derivative,
and other factors.

There is little theoretical justification for assumption that relative price
changes are normal, since the hypotheses for the Central Limit Theorem (CLT)
do not apply directly to prices, as the latter evolve through a function of
supply and demand. To be more specific, CLT involves the mean and variance of
a large number of independent, identically distributed random variables.
Prices are not in themselves independent random variables, but rather evolve
through a process involving supply and demand. Moreover, supply and demand
consist of a large number of independent agents placing buy and sell orders.
Each buyer (and likewise, each seller) can be deemed to be a random variable,
so that a large number of buyers can be reasonably approximated as independent
and identically distributed (regardless of the particular distribution) so the
case for applying the CLT is much stronger. See further discussion in Appendix D.

In this continuum formulation, the price equation can be written in its
simplest form in terms of relative excess demand as%
\begin{equation}
\tau_{0}\frac{1}{P\left(  t\right)  }\frac{dP\left(  t\right)  }{dt}%
=\frac{\mathcal{D}\left(  t\right)  -\mathcal{S}\left(  t\right)
}{\mathcal{S}\left(  t\right)  } \label{dpc}%
\end{equation}
where supply, $\mathcal{S},$ and demand, $\mathcal{D},$ can vary due to any
arbitrary set of motivations. For example, they can depend on the price, $P$,
and the discount from an assessment of the true value of the asset. Here,
$\tau_{0}$ is a time constant that also incorporates a constant rate factor
that would multiply the right hand side. A key difference between the
classical finance $\left(  \ref{cs}\right)  $ and the asset flow approaches is
due to fact that $\left(  \ref{cs}\right)  $ assumes infinite arbitrage. The
assumption is that there is always capital that can take advantage of
mispricing of assets. In this way the deviation from realistic value will be
small and random, and the random terms might be normal. In $\left(
\ref{dpc}\right)  $ the supply and demand can arise from various motivations
and strategies including undervaluation and price trend. A derivation and
discussion of this equation appear in $\cite{CC}$. See also modeling of
supply/demand in $\cite{HQ},$ $\cite{HGH}.$

While the discrete price change mechanism may be more complicated that the
models above indicate, one has to distinguish between a smoothly trading
asset, such as a large capitialization stock, and an asset that trades in
discrete quantities that cannot be approximated well by an equation such as
$\left(  \ref{dpc}\right)  $. Thus, $\left(  \ref{nonlin discrete p}\right)  $
may be a coarse approximation to a thinly traded asset, but a continuum limit
such as $\left(  \ref{dpc}\right)  $ that arises from an averaging process is
not likely to be sensitive to the infinitesimal perturbations of the supply
and demand.

Note that in the price equation $\left(  \ref{dpc}\right)  $, supply and
demand are not on an equal footing. This can be remedied by replacing the
right hand side by a function $G\left(  \mathcal{D}/\mathcal{S}\right)  $ with
particular properties (see Section 2.2 of $\cite{CC1}$), such as the function
$G\left(  u\right)  =u-u^{-1}.$ In other words, the simplest model, $\left(
\ref{dpc}\right)  $, may not be the one that corresponds to the empirical
evidence for price change for a particular asset market. Thus, we assume the
more general model%
\begin{equation}
\tau_{0}\frac{1}{P\left(  t\right)  }\frac{dP\left(  t\right)  }{dt}=G\left(
\frac{\mathcal{D}\left(  t\right)  }{\mathcal{S}\left(  t\right)  }\right)
\label{dpg}%
\end{equation}
with $G:\mathbb{R\rightarrow R}$ a differentiable function\footnote{While
differentiability is a desirable feature for studying dynamics, continuity
suffices for the tail behavior results in this paper.} that satisfies such
that $G^{\prime}>0$ and $G\left(  1\right)  =0,$ and possibly other conditions.

We assume that price changes satisfy $\left(  \ref{dpc}\right)  $ where
$\mathcal{D}$ and $\mathcal{S}$ are given by terms that include both Brownian
process, $W\left(  t\right)  $, i.e., randomness with normal distribution, as
well as jump discontinuities through a compound Poisson distribution.

\bigskip

We consider equation $\left(  \ref{dpc}\right)  $ for arbitrary correlation,
$\rho,$ between supply and demand, i.e., the most general case. There is some
subtlety to the issue of correlation, since the correlation can arise from
both the supply and demand conditioned on the number of jumps or shock in
each, as well as the correlation between the number of jumps (see Appendix C).
The special case $\rho=-1$ (anti-correlation) involves the assumption that
random terms that lead to higher demand also lead to lower supply with the
same magnitude, and that the number of jumps is identical. In this limiting
case one can use the exact density for a quotient using a modification of
\cite{CC} in the Appendix A. We consider all cases, and prove that the
following asymptotic relationship holds for each. For any $q>0,$ the function
obtained by smoothing \cite{E1} of the continuous function
\[
G\left(  x\right)  =sgn\left(  x-x^{-1}\right)  \left\vert x-x^{-1}\right\vert
^{1/q},
\]
for example, corresponds to a distribution that decays as $F\left(  x\right)
\sim\left\vert x\right\vert ^{-q}.$ The values $q\in\left[  3,5\right]  $
corresponds the mid-range of the empirical results \cite{GGP}, \cite{G}, as
discussed in the Conclusion.

While the results involve functions $G$ that are symmetric in supply and
demand, one can similarly consider asymmetric functions (as determined by
empirical data) whose properties will be manifested in the decay exponent
analogously. In this paper we also consider the density conditioned on
positive $\mathcal{D}$ and $\mathcal{S},$ in addition to unrestricted
$\mathcal{D}$ and $\mathcal{S}$. The conditioning does not alter the tail
exponent of the density.

In summary, a price dynamics equation based on excess demand yields a $q$
power law decay of the distribution function, whether or not there are jump
discontinuities, complementing the earlier result $\cite{CC1}$ that a quotient
of normals (without the Poisson jump terms) has a density with a power law
decay. Furthermore, there is a simple link between the power, $q^{-1},$ in the
deterministic price adjustment function, $G,$ above, and decay exponent as
indicated above.

There have been a variety of explanations for fat tails, such as the placement
of large orders. Related to this is the perspective that fat tails are a
consequence of jumps in prices. However, the placement of a large order has
the direct effect of a jump in supply or demand; the change is price is a
consequence. We model the changes in supply and demand that are responsible
for changes in price and show that the tail of the distribution has the same
power exponent with or without jump discontinuities.

Other approaches to modeling of prices and supply/demand and stochasticity
include $\cite{J},$ $\cite{CDC}.$ Research has also provided simulations and
theoretical models directly on prices (see, for example, \cite{RCF}, \cite{BH}).

From a mathematical perspective, the results obtained below are quite general.
In many physical or biological models there are two processes that have
Brownian motion and jump components (i.e., Levy processes of jump-diffusion
type $\cite{CT};$ see also \cite{RSS} for estimation and simulation using
multi-variate Levy processes) that are in some type of competition, so the
pertinent quantity is the difference between the two. However, as in the case
of supply/demand, it is generally the relative difference that one needs to
consider just to make the units balance. Thus, one has a quotient (with a
shift of a constant, $-1)$ of such Levy processes as a consequence. \ Hence,
the analysis below applies to all such problems.

For maximum generality in a variety of applications, we consider quotients in
which the numerator and denominator can be either (i) of either sign, or (ii)
are conditioned to be positive. The issue of the sign of demand or supply in
economics is considered in Appendix E.

\bigskip

\textbf{2. The density of the quotient of diffusion-jump processes with
arbitrary correlation and independent number of jumps. } We consider
stochastic processes consisting of Brownian plus a compound Poisson process
(see Cont and Tankov $\cite{CT}$ p. 111). \ Let $\mathcal{S}=X_{t}%
^{\mathcal{S}},$ $\mathcal{D}=X_{t}^{\mathcal{D}}$ be the stochastic supply
and demand\footnote{Note that we assume that both the supply and demand are
Levy jump diffusion processes, which, like a simple normal distribution, can
be negative. This issue is discussed in the Appendix.} at time $t,$ with
$\Delta X_{t}^{\mathcal{S},\mathcal{D}}:=X_{t+\Delta t}^{\mathcal{S}%
,\mathcal{D}}-X_{t}^{\mathcal{S},\mathcal{D}}$ for some finite interval,
$\Delta t.$ During a time interval $\left(  t,t+\Delta t\right)  $ one expects
that the deterministic contribution to the supply and demand will be
proportional to $\Delta t,$ i.e., $\Delta X_{t}^{\mathcal{S},\mathcal{D}%
}=\gamma_{\mathcal{S},\mathcal{D}}\Delta t$ (where $\gamma_{\mathcal{S}}$ and
$\gamma_{\mathcal{D}}$ are deterministic and independent of $\Delta t$) if
there is no randomness. We let $D:=\gamma_{\mathcal{D}}\Delta t$ and
$S:=\gamma_{\mathcal{S}}\Delta t.$ Note that $\gamma_{S}$ and $\gamma_{D}$ can
be regarded as the expected values of supply and demand (per unit time) and
can depend on the asset's price and price trend, and other factors that would
be coupled into the equations.

The stochastic supply and demand terms, $X_{t}^{\mathcal{S}},$ $X_{t}%
^{\mathcal{D}},$ incorporate a normal, i.e., Brownian motion, as well as a
compound Poisson process. In addition to the deterministic term and Brownian
motion, we allow for jumps whose timing is given by a Poisson process, while
the jump distribution itself is normal. Let $\Delta N_{t}$ be the number of
jumps in the demand during $\left(  t,t+\Delta t\right)  ,$ so that the
probability that there are $k$ jumps in an interval $\Delta t,$ is given by
the Poisson distribution,
\begin{equation}
\mathbb{P}\left(  \left\{  \Delta N_{t}=k\right\}  \right)  =e^{-\lambda
_{1}\Delta t}\frac{\left(  \lambda_{1}\Delta t\right)  ^{k}}{k!} \label{Pois}%
\end{equation}
and likewise for the supply, with $\tilde{N}_{t}$ similarly defined with
parameter $\lambda_{2}$.

A key point here is that $N_{t}$ and $\tilde{N}_{t}$ are independent. However,
the correlation between the numerator and denominator given $k_{1}$ jumps in
the numerator and $k_{2}$ in the denominator will be a fixed $\rho\in\left(
-1,1\right)  .$ In Section 5, we will consider a bivariate Poisson process
whereby the number of jumps $\Delta N_{t}$ and $\Delta\tilde{N}_{t}$ in the
demand and supply is arbitrarily correlated.

The jump processes in the numerator and denominator are defined by
\[
\sum_{i=N_{t}}^{N_{t+\Delta t}}Y_{i}\text{ with }Y_{i}\sim\mathcal{N}\left(
\mu_{1},\sigma_{1}^{2}\right)  ,\ \ \sum_{i=N_{t}}^{\tilde{N}_{t+\Delta t}%
}\tilde{Y}_{i}\text{ \ with }\tilde{Y}_{i}\sim\mathcal{N}\left(  \mu
_{2},\sigma_{2}^{2}\right)  .
\]
Also, let $W_{t}$ and $\tilde{W}_{t}$ denote Brownian motion, $\mathcal{N}%
\left(  0,\Delta t\right)  ,$ for the demand and supply, respectively. Thus
the ratio\footnote{The set of points on which the denominator vanishes has
measure zero, and the finiteness of the integral representing the distribution
is established in the sequel.} of the change in supply and demand during
$\left(  t,t+\Delta t\right)  $ is given by
\begin{equation}
\frac{\Delta X_{t}^{\mathcal{D}}}{\Delta X_{t}^{\mathcal{S}}}=\frac{D\left(
1+\frac{\sigma_{01}}{2}\Delta W_{t}+\sum_{i=N_{t}}^{N_{t+\Delta t}}%
Y_{i}\right)  }{S\left(  1+\frac{\sigma_{02}}{2}\Delta\tilde{W}_{t}%
+\sum_{i=N_{t}}^{\tilde{N}_{t+\Delta t}}\tilde{Y}_{i}\right)  }.
\label{deltaX}%
\end{equation}
The numerator and denominator are known as Levy processes of jump-diffusion
type $\cite{CT}.$ We make the standard assumption that $\left\{  \Delta
W_{t},Y_{i}\right\}  $ are all mutually independent, as are \ $\left\{
\Delta\tilde{W}_{t},\tilde{Y}_{i}\right\}  $. Note that we do not need to
multiply the $Y_{i}$ sum by a coefficient since it can be incorporated into
$\mu$ and $\sigma_{i}^{2}.$ The correlation, $\rho,$ between the random events
entering the supply and demand will be specified below. Empirically, the
supply and demand are generally negatively correlated, and often close to
$\rho=-1,$ however, we allow for the complete range $-1<\rho<1.$ The case
$\rho=-1$ will be considered in Section 4 through an exact expression for this
special case of anti-correlation. The case $\rho=1$ is the trivial case in
which the randomness in the supply and demand cancel, leaving only
deterministic terms in $\left(  \ref{deltaX}\right)  $. Further discussion of
the correlation is presented in Appendix C, where we calculate the total
correlation, $\rho_{T},$ between supply and demand and show that $\left\vert
\rho_{T}\right\vert $ is bounded away from $1$ when the jumps in the supply
and demand are independent. In Section 6, we consider jumps that are not
dependent and can consider the full range of $\rho_{T}.$

In the mathematical analysis below, it will be convenient to incorporate the
deterministic terms, $D$ and $S,$ into the constants as we consider the ratio
$R_{a}:=$ $R_{1}/R_{2}=$ $\Delta X_{t}^{\mathcal{D}}/\Delta X_{t}%
^{\mathcal{S}}$. Thus, we write (noting that $\mu_{01}$ and $\mu_{02}$ will be
proportional to $\Delta t)$%
\begin{align}
R_{1}  &  :=\mu_{01}+\frac{\sigma_{01}}{2}\Delta W+\sum_{i=N_{t}}^{N_{t+\Delta
t}}Y_{i};\nonumber\\
R_{2}  &  :=\mu_{02}+\frac{\sigma_{02}}{2}\Delta\tilde{W}+\sum_{i=\tilde
{N}_{t}}^{\tilde{N}_{t+\Delta t}}\tilde{Y}_{i}. \label{R1R2a}%
\end{align}
Hence, the initial mathematical problems will be to determine the tail of the
densities of $\left(  i\right)  $ $R_{a}=R_{1}/R_{2};$ $\left(  ii\right)  $
$R_{b}:=R_{a}-R_{a}^{-1},$ and $\left(  iii\right)  $ $R_{c}:=s\left(
R_{b}\right)  $ where $s$ is an arbitrary function for the price dynamics equation.

\bigskip

\textbf{Remark.} In principle, all of the analysis below will be valid even if
we had $\mu_{0i}$ proportional to $\left(  \Delta t\right)  ^{p}$ for $p>0.$
In other words, obtaining the decay of the density is not contingent on having
a meaningful limit as $\Delta t\rightarrow0.$

An alternative that is used in Section 4 is to consider $\left(
\ref{deltaX}\right)  $ without the $D/S$ (which is deterministic) factor, and
then note that the exponent of the tail density is not altered upon
multiplying the random variable by the deterministic factor $D/S.$

We proceed by first determining the density of $R_{a}$ denoted $f_{a}$
conditioned on the number of jumps, $k_{i}$, for $R_{i},$ so that $N_{t}%
=k_{1}$ and $\tilde{N}_{t}=k_{2}$ are fixed. With this conditioning, we define
a related variable where the\ $k_{i}$ are fixed:
\begin{align}
\hat{R}_{1}\left(  k_{1}\right)   &  =\mu_{01}+\frac{\sigma_{01}}{2}\Delta
W+Y_{1}+...+Y_{k_{1}}\nonumber\\
\hat{R}_{2}(k_{2})  &  =\mu_{02}+\frac{\sigma_{02}}{2}\Delta\tilde{W}%
+\tilde{Y}_{1}+...+\tilde{Y}_{k_{2}}. \label{R1R2h}%
\end{align}
Since the $Y_{i}$ are all independent, identically distributed variables
(i.i.d.), and also independent of $\Delta W,$ and likewise for $\tilde{Y}_{i}%
$, we have from $\left(  \ref{R1R2a}\right)  $ and $\left(  \ref{R1R2h}%
\right)  $)%
\begin{align}
\hat{R}_{1}\left(  k_{1}\right)   &  \sim\mathcal{N}\left(  \mu_{01}+k_{1}%
\mu_{1},\left(  \frac{\sigma_{01}}{2}\right)  ^{2}\Delta t+k_{1}\sigma_{1}%
^{2}\Delta t\right)  ,\nonumber\\
\hat{R}_{2}(k_{2})  &  \sim\mathcal{N}\left(  \mu_{02}+k_{2}\mu_{2},\left(
\frac{\sigma_{02}}{2}\right)  ^{2}\Delta t+k_{2}\sigma_{2}^{2}\Delta t\right)
.
\end{align}
We abbreviate $\sigma_{R_{1}}^{2}:=\left(  \frac{\sigma_{01}}{2}\right)
^{2}+k_{1}\sigma_{1}^{2}\Delta t,$ \ $\sigma_{R_{2}}^{2}:=$ $\left(
\frac{\sigma_{02}}{2}\right)  ^{2}+k_{2}\sigma_{2}^{2\Delta t},$ $\mu_{R_{1}%
}:=\mu_{01}+k_{1}\mu_{1},$ $\mu_{R_{2}}:=\mu_{02}+k_{2}\mu_{2},$ yielding,
$R_{i}\sim\mathcal{N}\left(  \mu_{R_{i}},\sigma_{R_{i}}\right)  ,$ $i=1,2.$

At this point we can assign an arbitrary correlation $\rho\in\left(
-1,1\right)  $ between the demand and supply (with $k_{1}$ and $k_{2}$ jumps
or shocks) as follows. For each fixed pair $\left(  k_{1},k_{2}\right)  $ one
sets, for a given $\rho,$%
\begin{equation}
Cov\left(  \hat{R}_{1}\left(  k_{1}\right)  ,\hat{R}_{2}(k_{2})\right)
=\sigma_{12}\left(  k_{1},k_{2}\right)  :=\rho\sigma_{R_{1}}\left(
k_{1}\right)  \sigma_{R_{2}}\left(  k_{1}\right)  . \label{Cov}%
\end{equation}
Thus, $\rho$ is independent of $k_{1}$ and $k_{2},$ and has the prescribed
value. Moreover, $\rho$ is an arbitrary parameter -- that can be determined
empirically -- as the correlation between the random events entering into the
supply and demand.

Given two normals with prescribed means and variances, we can write down a
joint density by specifying the correlation or covariance. In particular, the
joint density of $R_{1},$ $R_{2}$ conditioned on $\Delta N_{t}=k_{1}$ and
$\Delta\tilde{N}_{t}=k_{2}$ (see \cite{TO} p. 7), is%
\begin{align}
f^{\left(  k_{1},k_{2}\right)  }\left(  x_{1},x_{2}\right)  :=  &  f\left(
x_{1},x_{2}~|\ \Delta N_{t}=k_{1},\Delta\tilde{N}_{t}=k_{2}\right) \nonumber\\
&  =\frac{1}{2\pi\sigma_{R_{1}}\sigma_{R_{2}}}\frac{1}{\sqrt{1-\rho^{2}}}%
\exp\left\{  -\frac{1}{2\left(  1-\rho^{2}\right)  }Q\left(  x_{1}%
,x_{2}\right)  \right\} \nonumber\\
Q\left(  x_{1},x_{2}\right)   &  :=\left(  \frac{x_{1}-\mu_{R_{1}}}%
{\sigma_{R_{1}}}\right)  ^{2}-2\rho\left(  \frac{x_{1}-\mu_{R_{1}}}%
{\sigma_{R_{1}}}\right)  \left(  \frac{x_{2}-\mu_{R_{2}}}{\sigma_{R_{2}}%
}\right) \nonumber\\
&  +\left(  \frac{x_{2}-\mu_{R_{2}}}{\sigma_{R_{2}}}\right)  ^{2}.
\label{fcond}%
\end{align}
The full density of $R_{1}$ and $R_{2}$ without the conditioning on $k_{i}$ is
then%
\begin{align}
f\left(  x_{1},x_{2}\right)   &  =\sum_{k_{1},k_{2}=0}^{\infty}\mathbb{P}%
\left(  \Delta N_{t}=k_{1},\Delta\tilde{N}_{t}=k_{2}\right)  \times
\label{joint1}\\
\ \ \ \ \  &  \ \ \ \ \ \ \ f\left(  x_{1},x_{2}~|\ \Delta N_{t}=k_{1}%
,\Delta\tilde{N}_{t}=k_{2}\right) \nonumber
\end{align}%
\begin{align}
f\left(  x_{1},x_{2}\right)   &  =\sum_{k_{1},k_{2}=0}^{\infty}e^{-\lambda
_{1}\Delta t}\frac{\left(  \lambda_{1}\Delta t\right)  ^{k_{1}}}{k_{1}%
!}e^{-\lambda_{2}\Delta t}\frac{\left(  \lambda_{2}\Delta t\right)  ^{k_{2}}%
}{k_{2}!}\times\label{joint2}\\
\ \ \ \ \  &  \ \ \ \ \ \ \ f\left(  x_{1},x_{2}~|\ \Delta N_{t}=k_{1}%
,\Delta\tilde{N}_{t}=k_{2}\right)  .\nonumber
\end{align}

\textbf{Notation.} For asymptotic relationships, we use the standard notation
that for $g:\mathbb{R\rightarrow}\mathbb{R}$ we write for some $j\in
\mathbb{N}$, the relations $g\left(  x\right)  =O\left(  x^{-j}\right)  $ or
$g\left(  x\right)  \sim x^{-j}$ if there exist $c_{1},c_{2},x_{0}%
\in\mathbb{R}^{+}$ such that
\[
c_{1}x^{-j}\leq g\left(  x\right)  \leq c_{2}x^{-j}\text{ \ for all }x\geq
x_{0}.
\]
We will use the notation $g\left(  x\right)  \simeq x^{p}$ if the difference
$\left\vert g\left(  x\right)  -x^{p}\right\vert $ is bounded by a constant
times the next power, $x^{p+1}.$

\bigskip

\textbf{2.1a } \textbf{The density, }$f_{a}$\textbf{, of }$\boldsymbol{R}%
_{a}.$ With this joint density, $f\left(  x_{1},x_{2}\right)  ,$ of $R_{1}$
and $R_{2}$ above, one can write the quotient density (\cite{MGB} p. 187) as%
\[
f_{R_{1}/R_{2}}\left(  w\right)  =\int_{-\infty}^{\infty}\left\vert
y\right\vert f\left(  wy,y\right)  dy.
\]
For brevity let $c_{1}:=\left(  2\pi\sigma_{R_{1}}\sigma_{R_{2}}\sqrt
{1-\rho^{2}}\right)  ^{-1}$ and $c_{2}:=\left(  2\left(  1-\rho^{2}\right)
\right)  ^{-1}$ and write%
\begin{align}
f_{R_{1}/R_{2}}\left(  w\right)   &  =c_{1}\int_{-\infty}^{\infty}dy\left\vert
y\right\vert \sum_{k_{1},k_{2}=0}^{\infty}e^{-\lambda_{1}\Delta t}%
\frac{\left(  \lambda_{1}\Delta t\right)  ^{k_{1}}}{k_{1}!}e^{-\lambda
_{2}\Delta t}\frac{\left(  \lambda_{2}\Delta t\right)  ^{k_{2}}}{k_{2}!}%
\times\label{quotient}\\
&  \ \ \ \ \exp\left\{  -c_{2}\left[  \left(  \frac{wy-\mu_{R_{1}}}%
{\sigma_{R_{1}}}\right)  ^{2}-2\rho\left(  \frac{wy-\mu_{R_{1}}}{\sigma
_{R_{1}}}\right)  \left(  \frac{y-\mu_{R_{2}}}{\sigma_{R_{2}}}\right)
+\left(  \frac{y-\mu_{R_{2}}}{\sigma_{R_{2}}}\right)  ^{2}\right]  \right\}
.\nonumber
\end{align}

\bigskip

We denote
\[
I\left(  w\right)  :=\int_{-\infty}^{\infty}dy\left\vert y\right\vert
\exp\left\{  -c_{2}\left[  \left(  \frac{wy-\mu_{R_{1}}}{\sigma_{R_{1}}%
}\right)  ^{2}-2\rho\left(  \frac{wy-\mu_{R_{1}}}{\sigma_{R_{1}}}\right)
\left(  \frac{y-\mu_{R_{2}}}{\sigma_{R_{2}}}\right)  +\left(  \frac
{y-\mu_{R_{2}}}{\sigma_{R_{2}}}\right)  ^{2}\right]  \right\}
\]
and approximate for large $\left\vert w\right\vert .$ Let $z:=y-\frac
{\mu_{R_{1}}}{w}$ so $h\left(  z\right)  :=-\frac{c_{2}}{\sigma_{R_{1}}^{2}%
}z^{2}$ and
\begin{align*}
\left(  \frac{wy-\mu_{R_{1}}}{\sigma_{R_{1}}}\right)  \left(  \frac
{y-\mu_{R_{2}}}{\sigma_{R_{2}}}\right)   &  =\frac{wz\left(  z+\frac
{\mu_{R_{1}}}{w}-\mu_{R2}\right)  }{\sigma_{R_{1}}\sigma_{R_{2}}},\\
\left(  \frac{y-\mu_{R_{2}}}{\sigma_{R_{2}}}\right)  ^{2}  &  =\frac{\left(
z+\frac{\mu_{R_{1}}}{w}-\mu_{R2}\right)  ^{2}}{\sigma_{R_{2}}^{2}}.
\end{align*}
Using these identities we can write $I\left(  w\right)  $ as%
\[
I\left(  w\right)  =\int_{-\infty}^{\infty}dze^{w^{2}h\left(  z\right)
}g\left(  z;w\right)
\]
with
\[
g\left(  z;w\right)  :=\left\vert z+\frac{\mu_{R_{1}}}{w}\right\vert
\exp\left\{  2c_{2}\rho\frac{wz\left(  z+\frac{\mu_{R_{1}}}{w}-\mu_{R_{2}%
}\right)  }{\sigma_{R_{1}}\sigma_{R_{2}}}-c_{2}\frac{\left(  z+\frac
{\mu_{R_{1}}}{w}-\mu_{R2}\right)  ^{2}}{\sigma_{R_{2}}^{2}}\right\}  .
\]
We can also write, using
\begin{equation}
p_{k_{i}}:=e^{-\lambda_{i}\Delta t}\frac{\left(  \lambda_{i}\Delta t\right)
^{k_{i}}}{k_{i}!}, \label{Pabv}%
\end{equation}
the identity%
\begin{align}
f_{R_{1}/R_{2}}\left(  w\right)   &  =\sum_{k_{1},k_{2}=0}^{\infty}p_{k_{1}%
}p_{k_{2}}c_{1}\int_{-\infty}^{\infty}dze^{w^{2}h\left(  z\right)  }g\left(
z;w\right) \nonumber\\
&  =\sum_{k_{1},k_{2}=0}^{\infty}p_{k_{1}}p_{k_{2}}c_{1}I\left(  w\right)  .
\label{fw}%
\end{align}
The Laplace integral approximation\ (\cite{E} p. 36) then yields%
\begin{align}
I\left(  w\right)   &  \tilde{=}g\left(  0;w\right)  \left(  \frac{-2\pi
}{w^{2}h^{\prime\prime}\left(  0;w\right)  }\right)  ^{1/2}e^{w^{2}h\left(
0;w\right)  }\nonumber\\
&  +e^{w^{2}h\left(  0;w\right)  }O\left(  w^{-3}\right)  . \label{Iw}%
\end{align}
We compute $g\left(  0;w\right)  =\left\vert \frac{\mu_{R_{1}}}{w}\right\vert
\exp\left\{  -c_{2}\left(  \frac{\mu_{R_{2}}}{\sigma_{R_{2}}}\right)
^{2}\right\}  ,$ $h\left(  0\right)  =0,$ $h^{\prime\prime}\left(  z\right)
=-2\frac{c_{2}}{\sigma_{2}^{2}}.$ This yields%
\[
I\left(  w\right)  =\left\vert \frac{\mu_{R_{1}}}{w}\right\vert \exp\left\{
-c_{2}\left(  \frac{\mu_{R_{2}}}{\sigma_{R_{2}}}\right)  ^{2}\right\}
\pi^{1/2}\frac{\sigma_{R_{2}}}{c_{2}^{1/2}}\left\vert \frac{1}{w}\right\vert
.
\]
Combining the constants in $c_{1}I\left(  w\right)  $ we have
\[
\pi^{1/2}\frac{c_{1}}{c_{2}^{1/2}}\mu_{R_{1}}\sigma_{R_{2}}=\left(  \frac{\pi
}{2}\right)  ^{1/2}\frac{\mu_{R_{1}}}{\sigma_{R_{1}}}.
\]
Thus, we can write the density of the quotient $R_{1}/R_{2}$ as (for large
$\left\vert w\right\vert $ with the remainder term $O\left(  \left\vert
w\right\vert ^{-3}\right)  :$%
\begin{align}
f_{R_{1}/R_{2}}\left(  w\right)   &  \tilde{=}\sum_{k_{1},k_{2}=0}^{\infty
}e^{-\lambda_{1}\Delta t}\frac{\left(  \lambda_{1}\Delta t\right)  ^{k_{1}}%
}{k_{1}!}e^{-\lambda_{2}\Delta t}\frac{\left(  \lambda_{2}\Delta t\right)
^{k_{2}}}{k_{2}!}\label{R1R2}\\
&  \times\left(  \frac{\pi}{2}\right)  ^{1/2}\frac{\mu_{R_{1}}}{\sigma_{R_{1}%
}}\exp\left\{  -c_{2}\left(  \frac{\mu_{R_{2}}}{\sigma_{R_{2}}}\right)
^{2}\right\}  \frac{1}{\left\vert w\right\vert ^{2}}.\nonumber
\end{align}

\bigskip

\textbf{Remark}. The convergence of this sum is clear. In particular,
$\mu_{R_{i}}\ $and $\sigma_{R_{i}}$ depend on $k_{1}$ and $k_{2}$ (and so does
$c_{1}$ but not $c_{2}$), and one must consider these in terms of convergence
of the infinite series. Recall%
\[
\mu_{R_{1}}:=\mu_{01}+k_{1}\mu_{1},\ \ \mu_{R_{2}}:=\mu_{02}+k_{2}\mu_{2}%
\]%
\[
\sigma_{R_{1}}^{2}:=\left(  \frac{\sigma_{01}}{2}\right)  ^{2}+k_{1}\sigma
_{1}^{2},\ \ \sigma_{R_{2}}^{2}:=\left(  \frac{\sigma_{02}}{2}\right)
^{2}+k_{2}\sigma_{2}^{2}.
\]
Thus, for large $k_{1},$ $k_{2}$, one has $\mu_{R_{1}}/\sigma_{R_{1}}\sim
k_{1}^{1/2}$ and $\left(  \mu_{R_{2}}/\sigma_{R_{2}}\right)  ^{2}\sim k_{2}.$
Also, $\exp\left\{  -c_{2}\left(  \frac{\mu_{R_{2}}}{\sigma_{R_{2}}}\right)
^{2}\right\}  $ is less than $1.$ Thus, the end result is%

\begin{equation}
f_{R_{1}/R_{2}}\left(  w\right)  \sim\frac{e^{-\lambda_{1}\Delta t}%
e^{-\lambda_{2}\Delta t}}{\left\vert w\right\vert ^{2}}\sum_{k_{1},k_{2}%
=0}^{\infty}k_{1}^{1/2}\frac{\left(  \lambda_{1}\Delta t\right)  ^{k_{1}}%
}{k_{1}!}\frac{\left(  \lambda_{2}\Delta t\right)  ^{k_{2}}}{k_{2}!}<\infty.
\label{approx}%
\end{equation}

In other words, the factorials dominate the powers, and we have convergence
for all values of the parameters.

\bigskip

\textbf{2.2} \ \textbf{The density }$\boldsymbol{f}_{b}$\textbf{ of
}$\boldsymbol{R}_{b}\boldsymbol{:=r}\left(  \boldsymbol{R}_{a}\right)
\boldsymbol{:=R}_{a}\boldsymbol{-R}_{a}^{-1}$\textbf{.} Using the results
above, we now focus on the behavior of the density (for large, positive values
of the argument), $f_{b}$, for $R_{b}:=r\left(  R_{a}\right)  ,$ where
$R_{a}:=R_{1}/R_{2}$ with density $f_{a}$.

The function $y=r\left(  x\right)  :=x-x^{-1}$ has two smooth strictly
monotonic branches (for $x>0$ and $x<0$) as shown in Figure 1. The respective
inverses in the two regions are given by
\[
x_{\pm}=h_{\pm}\left(  y\right)  =\frac{1}{2}\left(  y\pm\sqrt{y^{2}%
+4}\right)  ,
\]%
\[
h_{\pm}^{\prime}\left(  y\right)  =\frac{1}{2}\left(  1\pm\left(
1+4/y^{2}\right)  ^{-1/2}\right)  .
\]

To calculate the density we first compute the large $\left\vert y\right\vert $
behavior for $h_{\pm}\left(  y\right)  $ and its derivatives $h_{\pm}^{\prime
}\left(  y\right)  $. For a given positive value of $y$, the two intersections
of $y$ with $r\left(  x\right)  $ will be denoted by $x_{+}^{>}$ on the right
half plane, and $x_{-}^{>}$ on the left half plane. Both values are on the
upper half plane as $y>0.$ See Figure 1.

For $y\gg1$ we have the approximations (with the error term being of the next
power of $\left\vert y\right\vert $)
\begin{equation}
x_{+}^{>}=h_{+}\left(  y\right)  \simeq y,\ \ x_{-}^{>}=h_{-}\left(  y\right)
\simeq\frac{1}{2}\left(  1-\left(  1+\frac{2}{y^{2}}\right)  \right)
=-\frac{1}{y}. \label{h}%
\end{equation}
In other words, when $y\gg1$ one has either $x\gg1$ (i.e., $x_{+}$ on the
right branch), or $-1\ll x<0$ (i.e., $x_{-}$on the left branch) as shown by
the intersections in Figure 1.

The derivatives for $y\gg1$ are approximated by%
\begin{equation}
h_{+}^{\prime}\left(  y\right)  \simeq1,\ \ h_{-}^{\prime}\left(  y\right)
\simeq\frac{1}{y^{2}}. \label{hpr}%
\end{equation}
The analogous computations for $y\ll-1$ yield, with the two intersections
$x_{\pm}^{<}$ on the right and left half planes (both in the lower half
plane):
\begin{align}
x_{+}^{<}  &  =h_{+}\left(  y\right)  \simeq\frac{1}{-y},\ \ x_{-}^{<}%
=h_{-}\left(  y\right)  \simeq y\nonumber\\
h_{+}^{\prime}\left(  y\right)   &  \simeq\frac{1}{y^{2}},\ \ \ h_{-}^{\prime
}\left(  y\right)  \simeq1. \label{hneg}%
\end{align}

In order to compute the density, $f_{b}\left(  y\right)  $, of $R_{b}$ we note
that each value of $y$ for $R_{b}$ can be attained in two ways represented by
the intersection of a constant value of $y$ with the two segments of $r\left(
x\right)  .$ Thus, one has%
\[
f_{b}\left(  y\right)  =f_{a}\left(  h_{+}\left(  y\right)  \right)
h_{+}^{\prime}\left(  y\right)  +f_{a}\left(  h_{-}\left(  y\right)  \right)
h_{-}^{\prime}\left(  y\right)  .
\]
For $y\gg1$ we use the relations above in $\left(  \ref{h}\right)  $ and
$\left(  \ref{hpr}\right)  $ together with the relation $f_{a}\left(
y\right)  \sim\left\vert y\right\vert ^{-2}$ to obtain%
\begin{equation}
f_{b}\left(  y\right)  \sim f_{a}\left(  y\right)  \cdot1+f_{a}\left(
-\frac{1}{y}\right)  \cdot\frac{1}{y^{2}}\sim\frac{1}{y^{2}}\ \ \ \text{for
}y\gg1. \label{fbpos}%
\end{equation}
Note that we have used, for $y\gg1,$ the relation $f_{a}\left(  -\frac{1}%
{y}\right)  \simeq1.$

Similarly, for $y\ll-1$ we have the relation%
\[
f_{b}\left(  y\right)  \sim f_{a}\left(  -\frac{1}{y}\right)  \frac{1}{y^{2}%
}+f_{a}\left(  y\right)  \cdot1\sim\frac{1}{y^{2}}.
\]
Thus we can conclude that $f_{b}\left(  y\right)  \sim\left\vert y\right\vert
^{-2}$ when $y\gg1.$

\bigskip

\textbf{2.3 } \textbf{The density for nonlinear functions in the price
equation.} We now consider a spectrum of nonlinear functions for the right
hand side of the price adjustment equation, and prove that the tail of the
density decays as a monomial, together with a calculation of the exponent.
Recalling that $R_{a}$ represents the quotient of demand and supply, each
modeled with a jump-diffusion process (i.e., Brownian motion together with a
compound Poisson process), we consider specifically, for $q>0$, the random
variable $R_{c}=G_{\varepsilon}\left(  R_{a}\right)  $ where $y=G_{\varepsilon
}\left(  x\right)  $ is a smoothing (see \cite{E1}) of each branch of%
\begin{equation}
G\left(  x\right)  =\left\{
\begin{array}
[c]{ccc}%
\left(  x-\frac{1}{x}\right)  ^{1/q} & if & x-\frac{1}{x}>0\\
-\left(  \frac{1}{x}-x\right)  ^{1/q} & if & x-\frac{1}{x}<0
\end{array}
\right.  . \label{G}%
\end{equation}
Note that the smoothing makes this differentiable at $x=\pm1.$ More general
functions $G$ can be considered provided they satisfy the conditions $G\left(
1\right)  =0,$ $G^{\prime}\left(  x\right)  >0$ for all $x\in\mathbb{R}%
\backslash\left\{  \pm1,0\right\}  $ and, if symmetry between $\mathcal{D}$
and $\mathcal{S}$ is imposed, $G\left(  x\right)  =-G\left(  x^{-1}\right)  $
(see \cite{CC1}).

We calculate the decay in the density in two different ways each of which is
useful for a general sets of functions replacing $G$.

\bigskip

\textbf{2.3a } \textbf{Calculating }$\boldsymbol{f}_{c}$\textbf{ through
}$\boldsymbol{f}_{b}.$ Given a positive real number $q$, we define a function
$s:\mathbb{R\rightarrow R}$ by
\[
s\left(  u\right)  :=\left\{
\begin{array}
[c]{ccc}%
u^{1/q} & if & u\geq0\\
-\left(  -u\right)  ^{1/q} & if & u<0
\end{array}
\right.  .
\]
and let $s_{\varepsilon}$ be the smoothing. Noting that $s_{\varepsilon}$ is
strictly monotonic, with $y=s_{\varepsilon}\left(  x\right)  $ and inverse,
$x=j_{\varepsilon}\left(  y\right)  ,$ which is a smoothing\footnote{The
$j_{\varepsilon}$ will not be given by the same formula as $s_{\varepsilon},$
but this is not relevant for our purpose.} of%
\[
j\left(  y\right)  =\left\{
\begin{array}
[c]{ccc}%
y^{q} & if & y\geq0\\
-\left(  -y\right)  ^{q} & if & y<0
\end{array}
\right.  .
\]
Recalling $R_{b}=r\left(  R_{a}\right)  $ where $r\left(  x\right)
:=x-x^{-1},$ the result $f_{b}\left(  y\right)  \sim\left\vert y\right\vert
^{-2}$ for $\left\vert y\right\vert \gg1$ and the density relation,%
\[
f_{c}\left(  y\right)  =f_{b}\left(  j_{\varepsilon}\left(  y\right)  \right)
j_{\varepsilon}^{\prime}\left(  y\right)  ,
\]
we have for $y\gg1$ the asymptotic relation,%
\[
f_{c}\left(  y\right)  \sim f_{b}\left(  y^{q}\right)  y^{q-1}\sim y^{-q-1},
\]
and for $y\ll-1$ the same end result,%
\[
f_{c}\left(  y\right)  \sim f_{b}\left(  -\left(  -y^{q}\right)  \right)
y^{q-1}\sim y^{-q-1}.
\]

\bigskip

\textbf{2.3b} \textbf{Calculating }$\boldsymbol{f}_{c}$\textbf{ from
}$\boldsymbol{f}_{a}$\textbf{ directly. }\ We are interested first in
$y=R_{c}\gg1,$ which can occur in two ways: $x\gg1,$ or $-1\ll x<0.$ We can
write the inverse of $G_{\varepsilon}$, denoted $H,$ in two continuous parts
that lie in the $x>0$ and $x<0$ half-planes. The superscripts $>$ and $<$
denote, respectively, the parts expression of $H$ that are above and below the
$x$ axis. Also, note that the regions $x-\frac{1}{x}>0$ correspond to $y>0$
for $y=G_{\varepsilon}\left(  x\right)  $ while $x-\frac{1}{x}<0$ corresponds
$y<0$ in $y=G\left(  x\right)  .$ The qualitative features for $G_{\varepsilon
}\left(  x\right)  $ are similar to that for $r\left(  x\right)  $ above, and
one can refer again to Figure 1.

In determining the density $f_{c}\left(  y\right)  $ for $y\gg1,$ we compute
first the inverse of $y=G_{\varepsilon}\left(  x\right)  =\left(  x-\frac
{1}{x}\right)  ^{1/q}$ in this region. Note that the roots of $x^{2}%
-y^{q}x-1=0$ are given by%
\[
H_{\pm}^{>}\left(  y\right)  :=x_{\pm}^{>}=\frac{y^{q}\pm\sqrt{y^{2q}+4}}{2}.
\]
The positive root corresponds to $x\gg1$ and is given by%
\[
H_{+}^{>}\left(  y\right)  =x_{+}^{>}=\frac{y^{q}}{2}\left(  1+\left(
1+\frac{4}{y^{2q}}\right)  ^{1/2}\right)  \tilde{=}\frac{y^{q}}{2}\left(
1+\left(  1+\frac{2}{y^{2q}}\right)  \right)  \simeq y^{q}.
\]

The negative root corresponds to the left branch of $G$ so $H_{-}^{>}\left(
y\right)  :=x_{-}^{>}$ is given by%
\begin{align*}
H_{-}^{>}\left(  y\right)   &  :=x_{-}^{>}=\frac{y^{q}}{2}\left(  1+\left(
1+\frac{4}{y^{2q}}\right)  ^{1/2}\right) \\
&  \simeq\frac{y^{q}}{2}\left(  1-\left(  1+\frac{2}{y^{2q}}\right)  \right)
\simeq-\frac{1}{y^{q}}.
\end{align*}
Computation of the derivatives yields%
\begin{align*}
\frac{d}{dy}H_{\pm}^{>}\left(  y\right)   &  =\frac{1}{2}\left[  qy^{q-1}%
\pm\frac{1}{2}\left(  y^{2q}+4\right)  ^{-1/2}2qy^{2q-1}\right] \\
&  =\frac{1}{2}qy^{q-1}\left(  1\pm\frac{y^{q}}{\left(  y^{2q}+4\right)
^{1/2}}\right)  ;\\
\frac{d}{dy}H_{+}^{>}  &  \sim y^{q-1}\text{ \ and \ }\frac{d}{dy}H_{-}%
^{>}\left(  y\right)  \sim y^{-q-1}%
\end{align*}

We can now compute the density as%
\[
f_{c}\left(  y\right)  =f_{a}\left(  H_{+}^{>}\left(  y\right)  \right)
\frac{d}{dy}H_{+}^{>}\left(  y\right)  +f_{a}\left(  H_{-}^{>}\left(
y\right)  \right)  \frac{d}{dy}H_{-}^{>}\left(  y\right)  ,
\]
yielding the approximation%
\[
f_{c}\left(  y\right)  \sim f_{a}\left(  y^{q}\right)  y^{q-1}+f_{a}\left(
-\frac{1}{y^{q}}\right)  y^{-q-1}.
\]
Recalling that $f_{a}\left(  -y^{-q}\right)  \tilde{=}f_{a}\left(  0\right)
\sim1,$ and $f_{a}\left(  z\right)  \sim z^{-2}$ we then have $f_{c}\left(
y\right)  \sim y^{-q-1}.$

\bigskip

We consider next the case $y\ll-1,$ which occurs in two ways: $0<x\ll1,$ or
$x\ll-1$ . These are the two intersections of a value of $y$ that is large and
negative corresponding to the right and left branches, respectively, and the
parts of $G_{\varepsilon}$ that lie below the $x-$axis.

On the negative parts of both branches we have the relation%
\[
y=-\left(  -x+\frac{1}{x}\right)  ^{1/q}%
\]
and thus $x^{2}+\left(  -y\right)  ^{q}-1=0,$ which has solutions
\begin{align*}
H_{\pm}^{<}\left(  y\right)   &  :=x_{\pm}^{<}=\frac{-\left(  -y\right)
^{q}\pm\sqrt{\left(  -y\right)  ^{2q}+4}}{2}\\
&  \simeq\frac{\left(  -y\right)  ^{q}}{2}\left(  -1\pm\left(  1+\frac
{2}{\left(  -y\right)  ^{2q}}\right)  \right)  .
\end{align*}
We have then%
\[
H_{+}^{<}\left(  y\right)  \sim\frac{1}{\left(  -y\right)  ^{q}}%
,\ \ \ \ \ \ H_{-}^{<}\sim-\left(  -y\right)  ^{q}.
\]
The derivatives are given by%
\[
\frac{d}{dy}H_{-}^{<}\left(  y\right)  \sim\left(  -y\right)  ^{q-1}%
,\ \ \ \frac{d}{dy}H_{+}^{<}\left(  y\right)  \sim\left(  -y\right)  ^{-q-1}.
\]

\bigskip

For $y\ll-1,$ we thereby compute the decay of the density as%
\begin{align*}
f_{c}\left(  y\right)   &  =f_{a}\left(  H_{+}^{<}\left(  y\right)  \right)
\frac{d}{dy}H_{+}^{<}\left(  y\right)  +f_{a}\left(  H_{-}^{<}\left(
y\right)  \right)  \frac{d}{dy}H_{-}^{<}\left(  y\right) \\
&  \sim f_{a}\left(  \frac{1}{\left(  -y\right)  ^{q}}\right)  \left(
-y\right)  ^{-q-1}+f_{a}\left(  -\left(  -y\right)  ^{q}\right)  \left(
-y\right)  ^{q-1}\\
&  \sim1\cdot\left(  -y\right)  ^{-q-1}+\left(  -y\right)  ^{-2q}\left(
-y\right)  ^{q-1}\sim\left(  -y\right)  ^{-q-1}.
\end{align*}

\bigskip

We have proven the following.

\bigskip

\textbf{Theorem 2.3.} Let $R_{a}:=R_{1}/R_{2}$ where $R_{1}=\mathcal{D},$
$R_{2}=\mathcal{S}$ are described by jump-diffusion processes (i.e., Brownian
motion plus a compound Poisson process) through $\left(  \ref{R1R2a}\right)
.$ Assume that the two Poisson processes, $\Delta N_{t}$ and $\Delta\tilde
{N}_{t}$ are independent Poisson processes with arbitrary parameters
$\lambda_{1}$ and $\lambda_{2}.$ Let $\rho\in\left(  -1,1\right)  $ be the
correlation between $R_{1}$ and $R_{2}$ conditioned on a fixed pair of jumps
$\left(  k_{1},k_{2}\right)  $. With $G_{\varepsilon}$ be defined via $\left(
\ref{G}\right)  ,$ the density $f_{c}$ of $R_{c}:=G_{\varepsilon}\left(
R_{a}\right)  $ satisfies the asymptotic relation
\[
f_{c}\left(  y\right)  \sim\left\vert y\right\vert ^{-q-1}\text{ \ \ for
}\left\vert y\right\vert \gg1.
\]

\bigskip

\textbf{3. The density functions conditioned on positivity.} We consider the
densities for functions of the ratio $R_{1}/R_{2}=\mathcal{D}/\mathcal{S}$ as
before except that supply, $\mathcal{S}$, and demand, $\mathcal{D},$ are
required to be positive random variables through conditioning.

\bigskip

\textbf{3.1 The quotient with conditioning}. The probability of $X/Y$
conditioned on $X>0$ and $Y>0$ is expressed as
\[
\mathbb{P}\left(  \frac{X}{Y}\leq u\ |\ X>0,Y>0\right)  =\frac{\mathbb{P}%
\left(  \frac{X}{Y}\leq u,X>0,Y>0\right)  }{\mathbb{P}\left(  X>0,Y>0\right)
}.
\]
Let $Q_{1}:=\mathbb{P}\left(  X>0,Y>0\right)  $ so that
\[
F_{X/Y}\left(  u\ |\ X>0,Y>0\right)  =Q_{1}^{-1}%
{\textstyle\iint\limits_{\substack{x/y\leq u,\\x>0,\ y>0}}}
f_{X,Y}\left(  x,y\right)  dxdy
\]
and using $w=x/y$ we have the conditional distribution and density as
\[
F_{X/Y}\left(  u\ |\ X>0,Y>0\right)  =Q_{1}^{-1}\int_{0}^{\infty}dy\int
_{0}^{u}yf_{X,Y}\left(  wy,y\right)  dw,
\]%
\[
f_{X/Y}\left(  u\ |\ X>0,Y>0\right)  =Q_{1}^{-1}\int_{0}^{\infty}%
yf_{X,Y}\left(  uy,y\right)  dy.
\]
In our case, the joint density $f_{X,Y}\left(  x_{1},x_{2}\right)  $ is given
by $f\left(  x_{1},x_{2}\right)  $ in $\left(  \ref{joint1}\right)  ,$
$\left(  \ref{joint2}\right)  .$

\bigskip

\textbf{3.2 Calculation of }$\boldsymbol{f}_{a}\left(  \boldsymbol{y\ |\ R}%
_{1}\boldsymbol{,R}_{2}\boldsymbol{>0}\right)  $\textbf{ for }$\left\vert
\boldsymbol{y}\right\vert \boldsymbol{\gg1.}$\textbf{ }The density of the
quotient $R_{a}:=R_{1}/R_{2}$ conditioned on $R_{1}$, $R_{2}>0$ is similar to
the calculation in Section 2.1, with the main difference being that the
integral $I\left(  w\right)  $ will involve only half of the interval, i.e.,
$\int_{0}^{\infty}yf_{X,Y}\left(  wy,y\right)  dy,$ versus $\int_{-\infty
}^{\infty}\left\vert y\right\vert f_{X,Y}\left(  wy,y\right)  dy.$ The
resulting asymptotics for large $\left\vert u\right\vert $ are the same
(except for constant factors). Hence, we have the same asymptotic relation,
\begin{equation}
f_{a}\left(  w\ |\ X>0,Y>0\right)  \sim\left\vert w\right\vert ^{-2}\text{
\ for ~}\left\vert w\right\vert \gg1. \label{fa}%
\end{equation}

We use this to obtain results on the decay of the density of $R_{c}%
=G_{\varepsilon}\left(  D/S\right)  $ under the condition that supply,
$\mathcal{S},$ and demand, $\mathcal{D}$ are both positive, utilizing two
different approaches. In the first approach we find the conditional density of
$R_{b}:=R_{a}-R_{a}^{-1}$ with $R_{a}:=R_{1}/R_{2},$ and then obtain the
conditional density of $f_{c}$ from $f_{b}$ which is obtained from $f_{a}.$ In
the second approach we obtain it from $f_{a}$ directly.

\bigskip

\textbf{3.3 \ Calculation of }$\boldsymbol{f}_{c}\left(  \boldsymbol{y\ |\ R}%
_{1}\boldsymbol{,R}_{2}\boldsymbol{>0}\right)  $\textbf{ through
}$\boldsymbol{f}_{b}\left(  \boldsymbol{y\ |\ R}_{1}\boldsymbol{,R}%
_{2}\boldsymbol{>0}\right)  .$ Defining $R_{b}:=r\left(  R_{a}\right)
=R_{a}-R_{a}^{-1}$, i.e., $y=r\left(  x\right)  =x-x^{-1}\ $and conditioning
on $R_{1}$ and $R_{2}$ positive, we have $R_{a}=R_{1}/R_{2}>0,$ so only the
right branch of $r\left(  x\right)  ,$ i.e., $x>0,$ is of interest. It has
inverse%
\[
h_{+}\left(  y\right)  :=x_{+}=\frac{1}{2}\left(  y+\sqrt{y^{2}+4}\right)  .
\]
As the conversion from $x$ to $y$ is single valued, the conditional density
thereby satisfies, for both $y\gg1$ and $y\ll-1,$ the identity
\begin{equation}
f_{b}\left(  y\ |\ R_{1},R_{2}>0\right)  =f_{a}\left(  h_{+}\left(  y\right)
\ |\ R_{1},R_{2}>0\right)  h_{+}^{\prime}\left(  y\right)  . \label{fbc}%
\end{equation}

Using the asymptotics obtained earlier, i.e.,$\left(  \ref{h}\right)  ,$
$\left(  \ref{hneg}\right)  ,$ we have for $y\gg1$ (so $x\gg1$) the asymptotic
relation,
\[
f_{b}\left(  y~|\ R_{1},R_{2}>0\right)  \sim f_{a}\left(  y\right)  \cdot1\sim
y^{-2}.
\]
and for $y\ll-1$ (so $-1\ll x_{+}<0$) the result%
\[
f_{b}\left(  y~|\ R_{1},R_{2}>0\right)  \sim f_{a}\left(  \frac{1}{-y}\right)
\frac{1}{y^{2}}\sim\frac{1}{y^{2}}.
\]

\bigskip

Thus, we conclude that the density of $R_{b}$ conditioned on positive $R_{1}$
and $R_{2}$ yields the same asymptotic relation
\begin{equation}
f_{b}\left(  y\ |\ R_{1},R_{2}>0\right)  \sim y^{-2},\ \ \text{ }\left\vert
y\right\vert >>1. \label{fb}%
\end{equation}

\bigskip

Next, we use this result to calculate $f_{c}\left(  y\ |\ R_{1},R_{2}%
>0\right)  .$ We define $s_{\varepsilon}$ in the same way as the smoothing of
\[
s\left(  u\right)  :=\left\{  \left(
\begin{array}
[c]{ccc}%
u^{1/q} & if & u\geq0\\
-\left(  -u\right)  ^{1/q} & if & u<0
\end{array}
\right)  \right.  .
\]
Note that $s_{\varepsilon}$ is strictly increasing and has an inverse
$h_{\varepsilon}$ with $h_{\varepsilon}\left(  y\right)  \sim y^{q}$ if
$\left\vert y\right\vert \gg1$ (i.e., the smoothing does not alter the
\ growth rate), and $h_{\varepsilon}^{\prime}\left(  y\right)  \sim3y^{2}.$

Thus, we have again, for $y\gg1$
\begin{align*}
f_{c}\left(  y\ |\ R_{1},R_{2}>0\right)   &  =f_{b}\left(  h_{\varepsilon
}\left(  y\right)  \ |\ R_{1},R_{2}>0\right)  h_{\varepsilon}^{\prime}\left(
y\right) \\
&  \sim\left(  y^{q}\right)  ^{-2}y^{q-1}.
\end{align*}
Similarly, we have for $y\ll-1,$%

\begin{align*}
f_{c}\left(  y\ |\ R_{1},R_{2}>0\right)   &  =f_{b}\left(  h_{\varepsilon
}\left(  y\right)  \ |\ R_{1},R_{2}>0\right)  h_{\varepsilon}^{\prime}\left(
y\right) \\
&  \sim\left(  -y\right)  ^{-2q}\left(  -y\right)  ^{q-1}=\left(  -y\right)
^{-q-1}.
\end{align*}

\textbf{3.4 \ Calculation of }$\boldsymbol{f}_{c}\ \left(
\boldsymbol{y\ |\ R}_{1}\boldsymbol{,R}_{2}\boldsymbol{>0}\right)  $\textbf{
through }$\boldsymbol{f}_{a}\left(  \boldsymbol{y\ |\ R}_{1}\boldsymbol{,R}%
_{2}\boldsymbol{>0}\right)  $ \textbf{directly}$.$ We proceed in the same way
as in Section 2.3b, where $R_{c}=G_{\varepsilon}\left(  R_{a}\right)  $ and
$y=G_{\varepsilon}\left(  x\right)  $ is a smoothing of
\[
G\left(  x\right)  =sign\left(  x-x^{-1}\right)  \left\vert x-x^{-1}%
\right\vert ^{1/q},\ \ \ \ \ q>0,
\]
Due to the conditioning on positive $R_{1},R_{2}$, we have $R_{a}=R_{1}%
/R_{2}>0$, so we only need to consider the positive $x$ branch of
$G_{\varepsilon}\left(  x\right)  .$ The inverse $H_{+}\left(  y\right)  $
(with $H_{+}^{>}\left(  y\right)  $ for $y>0$, $H_{+}^{<}\left(  y\right)  $
for $y<0$) is defined in the same \ way as in Section 2.3b, and $H_{-}$ is not relevant.

We have then for $y\gg1,$ the density relation (identical to Section 2.3b
except for the absence of $H_{-}$ term):
\[
f_{c}\left(  y\ |\ R_{1},R_{2}>0\right)  =f_{a}\left(  H_{+}^{>}\left(
y\right)  \ |\ R_{1},R_{2}>0\right)  \frac{d}{dy}H_{+}^{>}\left(  y\right)  ,
\]
yielding the approximation%
\[
f_{c}\left(  y\ |\ R_{1},R_{2}>0\right)  \sim f_{a}\left(  y^{q}%
\ |\ R_{1},R_{2}>0\right)  y^{q-1}+f_{a}\left(  -\frac{1}{y^{q}}%
\ |\ R_{1},R_{2}>0\right)  y^{-q-1}.
\]
Recalling that $f_{a}\left(  -y^{-q}|...\right)  \tilde{=}f_{a}\left(
0\right)  \sim1,$ and $f_{a}\left(  z|...\right)  \sim z^{-2}$ we then have
$f_{c}\left(  y|...\right)  \sim y^{-q-1}.$

Similarly, for $y\ll-1$, we compute the decay of the density as%
\begin{align*}
f_{c}\left(  y\ |\ R_{1},R_{2}>0\right)   &  =f_{a}\left(  H_{+}^{<}\left(
y\right)  |...\right)  \frac{d}{dy}H_{+}^{<}\left(  y\right) \\
&  \sim f_{a}\left(  \frac{1}{\left(  -y\right)  ^{q}}|...\right)  \left(
-y\right)  ^{-q-1}\sim\left(  -y\right)  ^{-q-1}.
\end{align*}
Thus, we have the conclusion that for $\left\vert y\right\vert \gg1,$ the
density satisfies the asymptotic relation%
\[
f_{c}\left(  y\ |\ R_{1},R_{2}>0\right)  \sim\left\vert y\right\vert ^{-q-1}.
\]
We have then the following result.

\bigskip

\textbf{Theorem 3.4.} Let $R_{a}:=R_{1}/R_{2}$ where $R_{1}=\mathcal{D},$
$R_{2}=\mathcal{S}$ are described by jump-diffusion processes (i.e., Brownian
motion plus a compound Poisson process) through $\left(  \ref{R1R2a}\right)
$. Assume that $\Delta N_{t}$ and $\Delta\tilde{N}_{t}$ are independent
Poisson processes with arbitrary parameters $\lambda_{1}$ and $\lambda_{2}.$
Let $\rho\in\left(  -1,1\right)  $ be the correlation between $R_{1}$ and
$R_{2}$ conditioned on a fixed pair of jumps $\left(  k_{1},k_{2}\right)  $.
With $G_{\varepsilon}$ defined via $\left(  \ref{G}\right)  ,$ the density
$f_{c}\left(  y\ |R_{1},R_{2}>0\right)  $ of $R_{c}:=G_{\varepsilon}\left(
R_{a}\right)  $ conditioned on positive $R_{1}$ and $R_{2}$ satisfies the
asymptotic relation
\[
f_{c}\left(  y\ |R_{1},R_{2}>0\right)  \sim\left\vert y\right\vert
^{-q-1}\text{ \ \ for }\left\vert y\right\vert \gg1.
\]

\textbf{Remark.} We conclude that the decay of the density $f_{c}$ is not
altered by the conditioning on positive supply and demand.

\bigskip

\textbf{4. \ The price function when supply and demand are anti-correlated
(}$\boldsymbol{\rho=-1}$\textbf{). \ }In this section, we consider the
remaining case in which the supply and demand (i.e., $R_{1}$ and $R_{2}$) have
correlation $\rho=-1.$ For this particular case one can also derive an exact
density from which one can analyze the $\Delta t\rightarrow0$ limits.

\bigskip

\textbf{4.1 \ The anti-correlated quotient of Levy processes of jump-diffusion
type. }We consider the quotient of demand, $\mathcal{D}=R_{1},$ and supply,
$\mathcal{S}=R_{2}$ within a small time interval, now for the remaining case
of $\rho=-1$, and$\boldsymbol{\ }$write this quotient more simply as
\begin{equation}
\frac{\mathcal{D}}{\mathcal{S}}=\frac{R_{1}}{R_{2}}=\frac{D\left(  \Delta
t+\frac{\sigma_{0}}{2}\Delta W_{t}+\sum_{i=N_{t}}^{N_{t+\Delta t}}%
Y_{i}\right)  }{S\left(  \Delta t-\frac{\sigma_{0}}{2}\Delta W_{t}%
-\sum_{i=N_{t}}^{N_{t+\Delta t}}Y_{i}\right)  }. \label{quot}%
\end{equation}
We assume still that $\left\{  \Delta W_{t},Y_{i}\right\}  $ are all mutually
independent, with $\Delta W_{t}\sim\mathcal{N}\left(  0,\Delta t\right)  $
again, and write $Y_{i}\sim\mathcal{N}\left(  \mu,\delta^{2}\right)  $ to
simplify notation.

The assumption of $\rho=-1$ in this limiting case implies that random events
that lead to a rise in demand will simultaneously give rise to a fall in
supply. Near the trading price anti-correlation is not far from a typical
situation, since incoming news that is negative (for example due to an
earnings downgrade) will cause a reassessment that impacts both the supply and
demand with similar (though not exactly equal) magnitudes. In practice, they
may have a strong negative correlation but greater than $-1.$

Note that $S$ and $D$ are the deterministic factors in supply and demand
(which we can regard as the expected values) while $\mathcal{S}$ and
$\mathcal{D}$ are the complete supply and demand functions with stochasticity.
We first define the quotient, $R_{a},$ without these factors:%
\[
R_{a}:=\frac{\Delta t+\frac{\sigma_{0}}{2}\Delta W_{t}+\sum_{i=N_{t}%
}^{N_{t+\Delta t}}Y_{i}}{\Delta t-\frac{\sigma_{0}}{2}\Delta W_{t}%
-\sum_{i=N_{t}}^{N_{t+\Delta t}}Y_{i}}.
\]
Since $Y_{i}\sim\mathcal{N}\left(  \mu,\delta^{2}\right)  ,$ if we examine
$R_{a},$ conditioned on fixed $\Delta N_{t}=N_{t+\Delta t}-N_{t}=k$, we have a
sum of $k+1$ normal variables plus a constant term $\Delta t$ in both the
numerator and denominator, which thus have the respective normal distributions%
\[
\mathcal{N}\left(  \Delta t\pm k\mu,\frac{\sigma_{0}^{2}}{4}\Delta
t+k\delta^{2}\right)  .
\]
An exact expression for a quotient of anti-correlated normal random variables,
such as $R_{a},$ was derived in Theorem 4.3 of\ $\cite{CC}$ under the
condition of anti-correlation. For quotient of several other distributions,
see $\cite{SN}$. If the numerator and denominator have distributions
$\mathcal{N}\left(  \mu_{1},\sigma_{1}^{2}\right)  $ and $\mathcal{N}\left(
\mu_{2},\sigma_{2}^{2}\right)  $ respectively, then the quotient has density
given by
\begin{equation}
f\left(  x\right)  =\frac{\mu_{1}\sigma_{2}+\mu_{2}\sigma_{1}}{\sqrt{2\pi}%
}\frac{\exp\left\{  -\frac{1}{2}\left(  \frac{\mu_{2}x-\mu_{1}}{\sigma
_{2}x+\sigma_{1}}\right)  ^{2}\right\}  \text{ }}{\left(  \sigma_{2}%
x+\sigma_{1}\right)  ^{2}}. \label{cc}%
\end{equation}
Now let $\sigma_{1}^{2}=\sigma_{2}^{2}=\sigma^{2}:=$ $\frac{\sigma_{0}^{2}}%
{4}\Delta t+k\delta^{2},$ $\mu_{1}:=\Delta t+k\mu$ and $\mu_{2}:=\Delta
t-k\mu$ so that this general formula yields%

\begin{equation}
f_{a}(x\ |\ \Delta N_{t}=k)=\frac{2\left(  \Delta t\right)  }{\sqrt{2\pi
}\sigma}\frac{\exp\left\{  -\frac{1}{2\sigma^{2}}\left(  \frac{\left(  \Delta
t\right)  \left(  x-1\right)  -k\mu\left(  x+1\right)  }{x+1}\right)
^{2}\right\}  }{\left(  x+1\right)  ^{2}}. \label{fq}%
\end{equation}

Next, we define $R_{A}=\frac{D}{S}R_{a},$ which has density%
\[
f_{A}(x\ |\ \Delta N_{t}=k)=\frac{f_{a}(\frac{x}{D/S}\ |\ \Delta N_{t}%
=k)}{D/S}.
\]
Substitution for $f_{a}$ yields
\begin{equation}
f_{k}\left(  x\right)  :=f_{A}(x\ |\ \Delta N_{t}=k)=\frac{2\Delta t}%
{\sqrt{2\pi}\sigma\frac{D}{S}}\frac{\exp\left\{  -\frac{1}{2\sigma^{2}}\left(
\frac{\left(  \Delta t\right)  \left(  \frac{x}{D/S}-1\right)  -k\mu\left(
\frac{x}{D/S}+1\right)  }{\frac{x}{D/S}+1}\right)  ^{2}\right\}  }{\left(
\frac{x}{D/S}+1\right)  ^{2}}. \label{fk}%
\end{equation}

\bigskip

\textbf{5. \ Asymptotics of the density.} \ We focus on the large $\left\vert
x\right\vert $ behavior of the density by first examining $\left(
\ref{fk}\right)  $ in the cases $k=0$ and $k\geq1$ separately.

$\left(  i\right)  $ For $k=0,$ recalling $\sigma^{2}:=$ $\frac{\sigma_{0}%
^{2}}{4}\Delta t+k\delta^{2}=\frac{\sigma_{0}^{2}}{4}\Delta t,$ we have simply%
\begin{align}
f_{0}\left(  x\right)   &  :=f_{A}(x\ |\ \Delta N_{t}=0)\label{f0}\\
&  =\frac{1}{\sqrt{2\pi}\frac{\sigma_{0}}{4\left(  \Delta t\right)  ^{\frac
{1}{2}}}}\frac{\frac{S}{D}}{\left(  \frac{x}{D/S}+1\right)  ^{2}}\exp\left\{
-\frac{1}{2}\frac{1}{\frac{\sigma_{0}^{2}}{4\Delta t}}\left(  \frac{\frac
{x}{D/S}-1}{\frac{x}{D/S}+1}\right)  ^{2}\right\}  .\nonumber
\end{align}
For $\left\vert x\right\vert \gg1$ this yields the expression
\begin{equation}
f_{0}\left(  x\right)  \sim\frac{1}{\sqrt{2\pi}\frac{\sigma_{0}}{4\left(
\Delta t\right)  ^{\frac{1}{2}}}}\frac{\frac{D}{S}}{x^{2}}\exp\left\{
-\frac{1}{2}\frac{1}{\frac{\sigma_{0}^{2}}{4\Delta t}}\right\}  . \label{f0x}%
\end{equation}

$\left(  ii\right)  $ For $k\geq1,$ \ and $\left\vert x\right\vert \gg1$ we
have similarly,%
\begin{equation}
f_{k}\left(  x\right)  \sim\frac{2\left(  \Delta t\right)  \frac{D}{S}}%
{\sqrt{2\pi}\sigma}\frac{\exp\left\{  -\frac{1}{2\sigma^{2}}\left[  \Delta
t-k\mu\right]  ^{2}\right\}  }{x^{2}}, \label{fkx}%
\end{equation}
so the last two expressions both have a power law decay, $x^{-2}.$

\bigskip\ 

Using the theorem of total probability, we now calculate the full density
(i.e., without conditioning on $k$)%
\begin{equation}
f_{A}\left(  x\right)  =\sum_{k=0}^{\infty}e^{-\lambda\Delta t}\frac{\left(
\lambda\Delta t\right)  ^{k}}{k!}\frac{2\Delta t}{\sqrt{2\pi}\sigma\frac{D}%
{S}}\frac{\exp\left\{  -\frac{1}{2\sigma^{2}}\left(  \frac{\mu_{2}\frac
{x}{D/S}-\mu_{1}}{\frac{x}{D/S}+1}\right)  ^{2}\right\}  }{\left(  \frac
{x}{D/S}+1\right)  ^{2}}. \label{fq1}%
\end{equation}

From the previously defined $f_{0},$ i.e., $\left(  \ref{f0}\right)  $ and
$f_{k}$ ($k\geq1$), one has
\begin{equation}
f_{A}\left(  x\right)  =e^{-\lambda\Delta t}f_{0}\left(  x\right)  +\sum
_{k=1}^{\infty}e^{-\lambda\Delta t}\frac{\left(  \lambda\Delta t\right)  ^{k}%
}{k!}f_{k}(x)\ . \label{fq1a}%
\end{equation}

Considering the $k\geq1$ terms separately, we have for $x\gg1,$%

\begin{equation}
\sum_{k=1}^{\infty}e^{-\lambda\Delta t}\frac{\left(  \lambda\Delta t\right)
^{k}}{k!}f_{k}(x)\ \sim\sum_{k=1}^{\infty}e^{-\lambda\Delta t}\frac{\left(
\lambda\Delta t\right)  ^{k}}{k!}\frac{2\left(  \Delta t\right)  \frac{D}{S}%
}{\sqrt{2\pi}\sigma}\frac{\exp\left\{  -\frac{1}{2\sigma^{2}}\left[  \Delta
t-k\mu\right]  ^{2}\right\}  }{x^{2}}. \label{sumx}%
\end{equation}

The series is clearly convergent so that the decay for large $\left\vert
x\right\vert $ is $O\left(  x^{-2}\right)  .$ Combining this with the similar
result for $f_{0}\left(  x\right)  $ yields the conclusion%

\begin{equation}
f_{A}\left(  x\right)  \sim x^{-2}\text{ \ \ \ as \ \ }\left\vert x\right\vert
\rightarrow\infty. \label{A}%
\end{equation}

The results of Section 2 for $\rho\in\left(  -1,1\right)  $ follow similarly
for the random variables $R_{b}:=R_{A}-R_{A}^{-1}$ and $R_{c}$%
:=$G_{\varepsilon}\left(  R_{A}\right)  $ where $G_{\varepsilon}$ is the
smoothing of $G$ defined by $\left(  \ref{G}\right)  .$

\bigskip

\textbf{Theorem 5.1.} Let $R_{A}$ be defined as the quotient of supply,
$\mathcal{S}$, and demand, $\mathcal{D},$ through $\left(  \ref{quot}\right)
$ with correlation $\rho=-1.$ With $G_{\varepsilon}$ be defined via $\left(
\ref{G}\right)  ,$ the density $f_{c}$ of $R_{c}:=G_{\varepsilon}\left(
R_{a}\right)  $ satisfies the asymptotic relation
\begin{equation}
f_{c}\left(  y\right)  \sim\left\vert y\right\vert ^{-q-1} \label{fc}%
\end{equation}
for $\left\vert y\right\vert \gg1.$

The density of $R_{c}$ conditioned on the numerator and denominator both being
positive also satisfies $\left(  \ref{fc}\right)  .$

\bigskip

Proof. The last assertion follows from the exact conditional density for the
quotient of anti-correlated normal random variables given in the Appendix A.
Since this expression differs from $\left(  \ref{cc}\right)  $ only by a
constant, all of the asymptotic relationships remain valid in this conditional
case. $///$

Next, we consider the full density under the conditions $\Delta t\ll1$ and
$x\gg1.$ The $f_{0}\left(  x\right)  $ remains the same while $f_{k}\left(
x\right)  $ for $k\geq1$ can be approximated using $\ \sigma^{2}=\frac
{\sigma_{0}^{2}}{4}\Delta t+k\delta^{2}\tilde{=}k\delta^{2}$ to yield the
following result for $\mu>0$. When $\mu=0$\textbf{ }with $\delta>0,$ i.e.,
$Y_{i}\sim\mathcal{N}\left(  0,\delta^{2}\right)  \ $we have no jumps in the
Poisson process, so the randomness arises only through the Brownian motion
which has been considered previously.

\bigskip

\textbf{Theorem 5.2 } Under the same conditions as Theorem 5.1, but with
$\Delta t\ll1$ in addition to $x\gg1,$ and $\mu>0,$ the density of $R_{A}$
\ is given by $\left(  \ref{fq1a}\right)  $ with $f_{0}$ and $f_{k}$ having
asymptotic behavior given by $\left(  \ref{f0x}\right)  $ and
\begin{equation}
f_{k}\left(  x\right)  \sim\frac{2\left(  \Delta t\right)  \frac{D}{S}}%
{\sqrt{2\pi}k^{1/2}\delta}\frac{\exp\left\{  -\frac{1}{2}\frac{k\mu^{2}%
}{\delta^{2}}\right\}  }{x^{2}} \label{dt}%
\end{equation}
for $\mu>0.$ Also, one has the bounds for sufficiently large $\left\vert
x\right\vert $%
\begin{align}
\frac{2}{\sqrt{2\pi}}\frac{\Delta t}{\delta}\frac{1}{x^{2}}\frac{D}%
{S}e^{-\lambda\Delta t}\frac{\lambda\Delta te^{-\frac{\mu^{2}}{2\delta^{2}}}%
}{2}  &  \leq\sum_{k=1}^{\infty}\frac{2\left(  \Delta t\right)  \frac{D}{S}%
}{\sqrt{2\pi}k^{1/2}\delta}\frac{\exp\left\{  -\frac{1}{2}\frac{k\mu^{2}%
}{\delta^{2}}\right\}  }{x^{2}}\label{b}\\
&  \leq\frac{2}{\sqrt{2\pi}}\frac{\Delta t}{\delta}\frac{1}{x^{2}}\frac{D}%
{S}e^{-\lambda\Delta t}\left[  \exp\left(  \lambda\Delta te^{-\frac{\mu^{2}%
}{2\delta^{2}}}\right)  -1\right]  .\nonumber
\end{align}

\bigskip

The proof is presented in Appendix B.

\bigskip

\textbf{Remark.} For nonlinear functions of $R_{A}$ one has similar
expressions and decay using the methods of the previous sections.

\bigskip

\textbf{Remark.} Adding to $\left(  \ref{dt}\right)  $ the $f_{0}\left(
x\right)  $ from $\left(  \ref{f0x}\right)  ,$ we see that asymptotic
expression for density $f_{A}\left(  x\right)  $ (for $\Delta t\ll1$ and
$x\gg1$) satisfies bounds obtained from $\left(  \ref{b}\right)  $ so that
$f_{A}\left(  x\right)  \sim$ $O\left(  1/x^{2}\right)  $ for $x\gg1.$

\bigskip

\textbf{6. The density with arbitrary correlation between the number of jumps
in the supply and demand. }In this section we allow for the possibility that
the number of jumps, $\Delta N_{t}$ in the demand (i.e., numerator) has an
arbitrary correlation, $\rho_{J},$ with the supply (i.e., denominator). For
positive reals $\lambda_{01},\ \lambda_{02,}\ \lambda_{12}$ and $k,l\in
\mathbb{N}$, we write the bivariate Poisson probability (see \cite{KK}) as%
\begin{equation}
p\left(  k,l\right)  :=\mathbb{P}\left(  \Delta N_{t}=k,\Delta\tilde{N}%
_{t}=l\right)  =\sum_{j=0}^{\min\left\{  k,l\right\}  }\frac{\lambda
_{01}^{k-j}\lambda_{02}^{l-j}\lambda_{12}^{j}e^{-\left(  \lambda_{01}%
+\lambda_{02}+\lambda_{12}\right)  }}{\left(  k-j\right)  !\left(  l-j\right)
!j!} \label{bivar}%
\end{equation}
where $\mathbb{E}\left[  \Delta N_{t}\right]  =Var\left[  \Delta N_{t}\right]
=\lambda_{01}+\lambda_{12}=:\lambda_{1},\ \ \mathbb{E}\left[  \Delta\tilde
{N}_{t}\right]  =Var\left[  \Delta\tilde{N}_{t}\right]  =\lambda_{02}%
+\lambda_{12}=:\lambda_{2}$. Summing over $l$ yields the individual
probabilities,
\[
\mathbb{P}\left(  \Delta N_{t}=k\right)  =\frac{\left(  \lambda_{01}%
+\lambda_{12}\right)  ^{k}}{k!}e^{-\left(  \lambda_{01}+\lambda_{12}\right)
},
\]
and similarly for $\mathbb{P}\left(  \Delta\tilde{N}_{t}=l\right)  .$ Also,
the covariance is given by $Cov\left[  \Delta N_{t},\Delta\tilde{N}%
_{t}\right]  =\lambda_{12}$ and, hence, the correlation by%
\begin{equation}
\rho^{\left(  P\right)  }:=\frac{\lambda_{12}}{\sqrt{\lambda_{01}+\lambda
_{12}}\sqrt{\lambda_{02}+\lambda_{12}}}. \label{corrP}%
\end{equation}
Thus the correlation in the number of jumps can be adjusted through the
parameters $\lambda_{01,}$ $\lambda_{02},$ and $\lambda_{12}$. The analysis of
Sections 2 and 3 can now be adapted for this general case by noting that the
joint density for the random variables $\hat{R}_{i}\left(  k_{i}\right)  $
defined by $\left(  \ref{R1R2h}\right)  $ is still given by $\left(
\ref{fcond}\right)  $ and $\left(  \ref{joint1}\right)  $. At this point we do
not assume independence of $\Delta N_{t}$ and $\Delta\tilde{N}_{t}$ but use
the joint probability density above, and $\left(  \ref{joint2}\right)  $ is
now replaced by
\begin{equation}
f\left(  x_{1},x_{2}\right)  =\sum_{k_{1},k_{2}=0}^{\infty}p\left(
k_{1},k_{2}\right)  \ f\left(  x_{1},x_{2}~|\ \Delta N_{t}=k_{1},\Delta
\tilde{N}_{t}=k_{2}\right)  .\nonumber
\end{equation}
and similarly, $\left(  \ref{fw}\right)  $ is identical except for the
$p\left(  k_{1},k_{2}\right)  $ substitution. The result $\left(
\ref{R1R2}\right)  $ is then
\begin{equation}
f_{R_{1}/R_{2}}\left(  w\right)  \tilde{=}\sum_{k_{1},k_{2}=0}^{\infty
}p\left(  k_{1},k_{2}\right)  \left(  \frac{\pi}{2}\right)  ^{1/2}\frac
{\mu_{R_{1}}}{\sigma_{R_{1}}}\exp\left\{  -c_{2}\left(  \frac{\mu_{R_{2}}%
}{\sigma_{R_{2}}}\right)  ^{2}\right\}  \frac{1}{\left\vert w\right\vert ^{2}%
}. \label{fw2}%
\end{equation}
One needs to show convergence of the sums above. Noting that $\mu_{R_{1}%
}/\sigma_{R_{1}}\sim k_{1}^{1/2}$ and the exponential is bounded by unity, one
can prove the boundedness of the series by noting that
\[
\sum_{k_{1},k_{2}=0}^{\infty}p\left(  k_{1},k_{2}\right)  k_{1}^{1/2}%
<\sum_{k_{1},k_{2}=0}^{\infty}p\left(  k_{1},k_{2}\right)  k_{1}%
=\mathbb{E}\left[  \Delta N_{t}\right]  =\lambda_{01}+\lambda_{12}<\infty.
\]

Hence, we can state the conclusion analogous to Theorem 2.3. In addition, the
analysis for $R_{1}$ and $R_{2}$ conditioned on positivity can be similarly
carried out as in Section 3, with the following result.

\bigskip

\textbf{Theorem 6.1.} Let $R_{a}:=R_{1}/R_{2}$ where $R_{1}=\mathcal{D},$
$R_{2}=\mathcal{S}$ are described by jump-diffusion processes (i.e., Brownian
motion plus a compound Poisson process) through $\left(  \ref{R1R2a}\right)
.$ Assume that the two Poisson processes, $\Delta N_{t}$ and $\Delta\tilde
{N}_{t}$ are bivariate Poisson with means $\lambda_{1}=\lambda_{01}%
+\lambda_{12}$ and $\lambda_{2}=\lambda_{02}+\lambda_{12}$ respectively, and
covariance $\lambda_{12}.$ Let $\rho\in\left(  -1,1\right)  $ be the
correlation between $R_{1}$ and $R_{2}$ conditioned on a fixed pair of jumps
$\left(  k_{1},k_{2}\right)  $. With $G_{\varepsilon}$ be defined via $\left(
\ref{G}\right)  ,$ the density $f_{c}$ of $R_{c}:=G_{\varepsilon}\left(
R_{a}\right)  $ satisfies the asymptotic relation
\[
f_{c}\left(  y\right)  \sim\left\vert y\right\vert ^{-q-1}\text{ \ \ for
}\left\vert y\right\vert \gg1.
\]
The density $f_{c}\left(  y\ |R_{1},R_{2}>0\right)  $ of $R_{c}%
:=G_{\varepsilon}\left(  R_{a}\right)  $ conditioned on positive $R_{1}$ and
$R_{2}$ also satisfies the asymptotic relation
\[
f_{c}\left(  y\ |R_{1},R_{2}>0\right)  \sim\left\vert y\right\vert
^{-q-1}\text{ \ \ for }\left\vert y\right\vert \gg1.
\]
The total correlation, $\rho_{T},$ as a function of the parameters is
calculated in the Appendix C.

\bigskip

Summarizing the results, we see that the issue of correlation between supply
and demand is complex. Given $k_{1}$ jumps or shocks for demand and $k_{2}$
for supply, we can impose a correlation, $\rho\in\left(  -1,1\right)  ,$
between supply and demand conditioned on $\left(  k_{1},k_{2}\right)  .$
However, the overall correlation (not conditioned on $k_{1}$ and $k_{2}$) for
demand and supply will have a smaller magnitude if the number of jumps are
independent, and is calculated in Appendix C.

In this section, we have considered the problem when the number of jumps in
the supply and demand, $\Delta N_{t}$ and $\Delta\tilde{N}_{t}$ has an
arbitrary correlation, $\rho^{\left(  P\right)  },$ given by $\left(
\ref{corrP}\right)  $ in which case one can obtain the full range $\left(
-1,1\right)  $ of correlation for the supply and demand, as stated in Theorem 6.1.

The limiting case in which the correlation between supply and demand is given
by $-1,$ i.e., anti-correlation, has been considered in Theorem 5.1.

\bigskip

\bigskip

\textbf{7. Conclusion. }We have shown that fat tails can be obtained solely by
modeling price adjustment through supply and demand, each of which are random
variables consisting of Brownian motion and shocks from a Poisson
distribution. This consistent with the empirical evidence for many asset
markets. The classical theory embodied in $\left(  \ref{cs}\right)  $ is based
partly on empirical observations near the mean (and not the tails) of price
changes. The fact that the normal distribution is a reasonable model near the
mean -- and is generally convenient mathematically -- does not provide any
evidence that the tail of density should behave as the normal does, i.e.,
$f\left(  x\right)  \sim e^{-x^{2}/2}.$ Prices evolve through re-balancing
supply and demand. In this paper we have assumed a model that embodies the
perspective of supply/demand without invoking the fine microstructure, and
show that the quotient of two Levy processes of jump-diffusion type has a tail
distribution that is a power law.

In essence, exponential decay in the classical finance models is a feature of
the assumptions made without compelling theoretical arguments. Prices are not
distributed by a random process; rather they are the consequence of changes in
supply and demand which have randomness. In fact, it is almost a tautology to
say that all price change is a direct consequence of changes in supply and
demand, which have randomness. If the supply and demand were deterministic,
there would be no randomness in price change. Aggregate supply and demand
involve the sum of a large number of decision makers so that the Central Limit
Theorem can be applied. The addition of a compound Poisson process to the
randomness incorporates the influx of a smaller number of large orders that
are manifested as jumps in supply or demand.

A natural way to model price change is through an excess demand model which
involves a quotient of supply and demand. The important mathematical idea is
that a quotient of two normals is not normal, though it resembles a normal
near the mean for a range of parameters. The tail of the density of a quotient
of normals has previously been shown to have a power law decay \cite{CC}. In
this paper, we showed that this persists even with the introduction of shocks
in the form of a Poisson distribution when there is an arbitrary correlation
in the number of shocks and an arbitrary total correlation between supply and demand.

Another important issue is that the price adjustment equation (\ref{dpc}) is
not necessarily linear. Hence, we consider appropriate nonlinear functions,
$G_{\varepsilon}$, $\left(  \ref{G}\right)  $\ that are characterized by an
exponent $q^{-1}$ for $q>0.$ The salient feature of this function is that
supply and demand enter on equal footing.

For this family of functions we have calculated the power of the decay
exponent for the density of the relative price change $P^{-1}\Delta P$, as
$f\left(  x\right)  \sim x^{-q-1}.$ The exponent is a function of the shape of
the price function, $G_{\varepsilon},$ so that there is a direct link between
the form of the price function and the decay exponent. Other possibilities for
the function $G_{\varepsilon}$ for supply and demand yields similar results.
Roughly speaking if the dependence of $G$ on the quotient of demand to supply
is characterized by the power $q^{-1}$, then the decay will be $f\left(
x\right)  \sim\left\vert x\right\vert ^{-\alpha}$ with $\alpha=1+q.$

Empirical studies of the tail of distributions for relative stock price
changes (denoted by $r$ below) have mainly shown (see \cite{G}, \cite{GGP},
\cite{RMF}, \cite{R}, \cite{BP}, \cite{MS} and references therein) that large
capitalization stocks have a decay rate given by $\zeta\tilde{=}3$ or at least
within a range $\zeta\in\left[  3,5\right]  $ where%
\[
P\left(  \left\{  \left\vert r\right\vert >x\right\}  \right)  \sim x^{-\zeta
},
\]
so that $f\left(  x\right)  \sim x^{-\zeta-1}.$ An early paper \cite{GM}
studied 40 million data points and hypothesized a $\zeta=3$ "universal cubic
law." Thus, we have $\alpha=q+1$, i.e., $\zeta=q.$ This would predict that the
$G_{\varepsilon}$ defined above should have $q^{-1}=$ $1/3$ corresponding to
$\alpha=4$ \ $\zeta=3.$ This means that if the nonlinear price function
$G_{\varepsilon}$ has a behavior that is somewhat less responsive than linear,
the implied decay will be within the empirical range. The analysis also
suggests that it would be useful to do empirical research on determining the
form of the function $G_{\varepsilon}$, as well as examining the correlation
between the changes in supply and demand to determine how close it is to
anti-correlation. Some studies (e.g., \cite{LS}) have indicated that the tail
exponent varies to some extent as one examines different countries and markets
such as emerging markets. While stock price movements have been studied
extensively, there is much less empirical work on the nature of the supply and
demand curves, particularly under extreme conditions. This would be a useful
empirical work that could open up a new avenue to understand rare events as
discussed below. This approach can be utilized not only for the tail but for
the entire distribution so volatility can be modeled in its entirety.

Our model is applicable to any price formation, and is not restricted to
financial instruments. In some markets for goods, the exponential tail may
hold \cite{KM}, \cite{AY}. Our analysis may provide a link between the tails
of financial assets and goods/services. In our model, this would be attained
by a function $G$ that is very slowly responding for large imbalance e.g.,
involving a logarithmic function. This would mean that as the supply/demand
balance becomes extreme, the relative price does not continue to change nearly
as much. It seems intuitively reasonable that the supply and demand curves for
goods/services have this feature, since there is usually no possibility of
resale. However, further research is needed on this issue.

In general, $G$ is a function to be calibrated. The essential point, however,
is that $G$ is a deterministic function, while $\mathcal{D}$ and $\mathcal{S}$
are Levy processes of jump-diffusion type. To simplify the discussion, let us
suppose that they are simply Gaussian. The central thesis is that price change
is essentially deterministic given a specific supply and demand. Suppose that
one is interested in estimating the probability of a $5\%$ drop in an asset.
In order to establish this probability directly from empirical data, one would
need a very large amount of data since it occurs infrequently. However, using
the formalism above, one would need many fewer observations that would
determine the value of the price drop as a function of the imbalance in supply
and demand. Once this connection is established, it is then a matter of
understanding the probability of that imbalance. The supply, $\mathcal{S},$
and demand, $\mathcal{D},$ are each Gaussian, with means and variances that
are easily estimated with a moderate amount of data (i.e., without necessarily
observing rare events) so we know the density of the normalized excess demand,
namely, $\left(  \mathcal{D-S}\right)  /\mathcal{S}$.

In summary, we can establish the deterministic connection between the
imbalance of supply/demand and the change in price through the function $G.$
Once this is done, we can estimate the means and variances of the supply and
demand individually. The mathematical results then yield the density of
$\left(  \mathcal{D-S}\right)  /\mathcal{S}$. With the approximation of $G$ we
then obtain the probability of a particular size price drop. In particlar, we
can estimate the distribution tail exponent, $\zeta=q.$

Note that the use of a smaller data set is not just a matter of convenience.
The issue is that as we use much larger sets, we are utilizing data that may
be irrelevant, for example, because it is an artifact of a different era. Data
sets of 252 trading days and 2520 correspond to one year and ten years,
respectively. In the year 2011, for example, the ten year period would include
the post-Internet bubble of 2001-2003, as well as the 2007-2008 housing
debacle and subsequent recovery. So data from only the period 2010-2011 would
be more pertinent in terms of estimating the probabilities of possible rare
events in 2012. Then using the data from 252 or 504 relevent trading days, one
can find the most extreme price changes and match them up with the $\left(
\mathcal{D-S}\right)  /\mathcal{S}$ values, so that the values of $G$ for
those arguments are set. This is a consequence of $G$ being deterministic.
From this data set of 252 or 504 we can reliably determine the mean and
variance of each of supply and demand, noting that these are Gaussian by
virtue of the Central Limit Theorem for independent agents. Using these
parameters, and the analysis of the quotient of Levy jump-diffusion processes,
we can calculate a more accurate probabilty for the extreme events that were
observed than would be possible by using just the raw frequency.

The problem of fat tails is part of the general issue of volatility in markets
that has been of great interest in recent years (see, e.g., \cite{FP}). The
methods discussed can be useful within the general setting of volatility problems.

The analysis of this paper has involved normal distributions through the
Brownian motion and the distribution of each jump in the Poisson process. An
open problem is to extend these results to a broad set of distributions. Using
results such as \cite{SN}, one can consider the quotient other distributions
for the supply and demand and analyze the tail behavior.

\bigskip

\textbf{Appendix A.} \textbf{The density of a quotient of anti-correlated
normal random variables with positive numerator and denominator.}

We obtain a result that is a modification of Theorem 4.3 of \cite{CC}. We
define, for $\mu_{1},\sigma_{1},\mu_{2},\sigma_{2}>0$ and $Z\sim
\mathcal{N}\left(  0,1\right)  $%
\[
R:=\frac{\mu_{1}+\sigma_{1}Z}{\mu_{2}-\sigma_{2}Z}%
\]
and wish to calculate (neglecting sets of measure zero)
\[
P_{+}:=\mathbb{P}\left\{  R\leq x\ |\ \mu_{1}+\sigma_{1}Z>0,\ \mu_{2}%
-\sigma_{2}Z>0\right\}  =\frac{\mathbb{P}\left\{  R\leq x,\ \mu_{1}+\sigma
_{1}Z,\ \mu_{2}-\sigma_{2}Z>0\right\}  }{\mathbb{P}\left\{  \ \mu_{1}%
+\sigma_{1}Z>0,\ \mu_{1}-\sigma_{1}Z>0\right\}  }%
\]
Let $Q:=\mathbb{P}\left\{  \ \mu_{1}+\sigma_{1}Z>0,\ \mu_{1}-\sigma
_{1}Z>0\right\}  $ so that $Q=\int_{-\mu_{1}/\sigma_{1}}^{\mu_{2}/\sigma_{2}%
}\left(  2\pi\right)  ^{-1/2}e^{-x^{2}/2}dx.$ We have then%
\[
QP_{+}=\mathbb{P}\left\{  \mu_{1}+\sigma_{1}Z\leq\left(  \mu_{2}-\sigma
_{2}Z\right)  x,\ \ \mu_{1}+\sigma_{1}Z>0,\ \mu_{1}-\sigma_{1}Z>0\right\}  .
\]
The three inequalities above reduce to $\left(  i\right)  $ $Z\leq\frac
{\mu_{2}x-\mu_{1}}{\sigma_{2}x+\sigma_{1}}$, $\left(  ii\right)  $ $Z\geq
-\mu_{1}/\sigma_{1},$ and $\left(  iii\right)  $ $Z<\mu_{2}/\sigma_{2}.$ Since
the inequality $\left(  iii\right)  $ is less restrictive than $\left(
i\right)  ,$ we can ignore $\left(  iii\right)  $ and write, with
$f_{Z}\left(  s\right)  =\left(  2\pi\right)  ^{-1/2}e^{-x^{2}/2}$ as the
density for $Z\sim\mathcal{N}\left(  0,1\right)  :$%
\[
QP_{+}=\int_{-\mu_{1}/\sigma_{1}}^{\frac{\mu_{2}x-\mu_{1}}{\sigma_{2}%
x+\sigma_{1}}}f_{Z}\left(  s\right)  ds.
\]
Differentiating this expression with respect to $x$ yields the result that we
can express in the form%
\[
f_{X/Y}\left(  x\ |X>0,\ Y>0\right)  =\frac{\mu_{1}\sigma_{2}+\mu_{2}%
\sigma_{1}}{\sqrt{2\pi}Q}\frac{e^{-\frac{1}{2}\left(  \frac{\mu_{2}x-\mu_{1}%
}{\sigma_{2}x+\sigma_{1}}\right)  }}{\left(  \sigma_{2}x+\sigma_{1}\right)
^{2}}.
\]

\bigskip\bigskip

\textbf{Appendix B. Proof of Theorem 5.2. \ }We examine bounds on these
asymptotic expressions for the full density if both $\Delta t\ll1$ and $x\gg1$
[i.e.,$\left(  \ref{fq1a}\right)  $ with $\left(  \ref{f0x}\right)  $ and
$\left(  \ref{dt}\right)  $] and obtain
\begin{align*}
\sum_{k=1}^{\infty}f_{k}\left(  x\right)  e^{-\lambda\Delta t}\frac{\left(
\lambda\Delta t\right)  ^{k}}{k!}  &  \sim\frac{2}{\sqrt{2\pi}}\frac{\Delta
t}{\delta}\frac{D}{S}\frac{1}{x^{2}}e^{-\lambda\Delta t}\sum_{k=1}^{\infty
}\frac{\left(  \lambda\Delta t\right)  ^{k}\exp\left\{  -\frac{1}{2}\frac
{k\mu^{2}}{\delta^{2}}\right\}  }{k^{1/2}k!}\\
&  =CA,
\end{align*}%
\[
C:=\frac{2}{\sqrt{2\pi}}\frac{\Delta t}{\delta}\frac{D}{S}\frac{1}{x^{2}%
}e^{-\lambda\Delta t},\ \ A:=\sum_{k=1}^{\infty}\frac{\left(  \lambda\Delta
t\right)  ^{k}\exp\left\{  -\frac{1}{2}\frac{k\mu^{2}}{\delta^{2}}\right\}
}{k^{1/2}k!}.
\]
Upon defining%
\[
A_{1}:=\sum_{k=1}^{\infty}\frac{\left(  \lambda\Delta t\right)  ^{k}%
\exp\left\{  -\frac{1}{2}\frac{k\mu^{2}}{\delta^{2}}\right\}  }{k!}%
,\ \ \ A_{2}:=\sum_{k=1}^{\infty}\frac{\left(  \lambda\Delta t\right)
^{k}\exp\left\{  -\frac{1}{2}\frac{k\mu^{2}}{\delta^{2}}\right\}  }{\left(
k+1\right)  !}.
\]
we have%
\begin{equation}
A_{2}\leq A\leq A_{1}.
\end{equation}
We let $a:=\lambda\Delta te^{-\frac{1}{2}\frac{\mu^{2}}{\delta^{2}}}$ so
\begin{align}
A_{1}  &  =e^{a}-1\\
A_{2}  &  =\frac{1}{a}\left(  e^{a}-1-a\right)  \geq a/2.\nonumber
\end{align}
satisfy the bounds%
\[
CA_{2}\leq CA\leq CA_{1}.
\]%
\begin{equation}
C\frac{a}{2}\leq C\frac{1}{a}\left(  e^{a}-1-a\right)  =CA_{2}\leq CA\leq
CA_{1}\leq C\left(  e^{a}-1\right)  ,\ i.e.\text{,}%
\end{equation}
yielding $\left(  \ref{b}\right)  $.

Noting also that the $f_{0}\left(  x\right)  $ term cannot be simplified
further beyond $\left(  \ref{f0x}\right)  $ for $\Delta t\ll1$ we see from
$\left(  \ref{fq1a}\right)  $ that all terms decay as $x^{-2}$ for $x>>1.$

\bigskip

\textbf{Appendix C.} \textbf{Analysis of supply/demand correlation.} In
Section 2 we considered an arbitrary correlation, $\rho$, between supply and
demand when there are $k_{1}$ jumps or shocks in demand, and $k_{2}$ in supply
during a small time interval, $\Delta t,$ determined by independent Poisson
processes, though the magnitude of the jumps in supply and demand are
correlated. The independence of the Poisson processes implies an upper bound
to the magnitude of the correlation between the supply and demand. In Section
6 we utilized a bivariate Poisson distribution for the jumps, which means that
the full range $\rho\in\left(  -1,1\right)  $ can be attained. Both of these
assertions will be established, together with an exact calculation in this Appendix.

We start with $R_{1}$ and $R_{2}$ defined as earlier by%
\[
R_{1}=\mu_{01}+\frac{\sigma_{01}}{2}\Delta W+\sum_{j=0}^{\Delta N_{t}}%
Y_{j},\ \ R_{2}=\mu_{02}+\frac{\sigma_{02}}{2}\Delta\tilde{W}+\sum
_{j=0}^{\Delta N_{t}}\tilde{Y}_{j}%
\]
where $Y\sim\mathcal{N}\left(  0,\sigma_{1}\right)  $ and $\tilde{Y}%
\sim\mathcal{N}\left(  0,\sigma_{2}\right)  ,$ and we have set $\mu_{1}%
=\mu_{2}=0$ so that the drift arises only from the first term, implying
$\mathbb{E}R_{1}=\mu_{01},$ $\mathbb{E}R_{2}=\mu_{02}.$ The variables $\hat
{R}_{1}\left(  k_{1}\right)  $ and $\hat{R}_{2}\left(  k_{2}\right)  $ are the
corresponding values when $\Delta N_{t}=k_{1}$ and $\Delta\tilde{N}_{t}%
=k_{2},$ namely, $\left(  \ref{R1R2h}\right)  .$

We calculate, using the notation $\mathbb{P}\left\{  \Delta N_{t}%
=k_{1},\ \Delta\tilde{N}_{t}=k_{2}\right\}  =:p\left(  k_{1},k_{2}\right)  $,
the identity,%
\begin{align*}
VarR_{1}  &  =\mathbb{E}\left[  \left(  R_{1}-\mathbb{E}R_{1}\right)
^{2}\right]  =\mathbb{E}\left[  \left(  \frac{\sigma_{01}}{2}\Delta
W+\sum_{j=0}^{\Delta N_{t}}Y_{j}\right)  ^{2}\right] \\
&  =\sum_{k_{1},k_{2}=0}^{\infty}\mathbb{E}\left[  \left(  \frac{\sigma_{01}%
}{2}\Delta W+\sum_{j=0}^{\Delta N_{t}}Y_{j}\right)  ^{2}\ \ |\ \Delta
N_{t}=k_{1},\ \Delta\tilde{N}_{t}=k_{2}\right]  p\left(  k_{1},k_{2}\right) \\
&  =\sum_{k_{1},k_{2}=0}^{\infty}\mathbb{E}\left[  \left(  \frac{\sigma_{01}%
}{2}\Delta W+\sum_{j=0}^{k_{1}}Y_{j}\right)  ^{2}\ \right]  p\left(
k_{1},k_{2}\right)  .
\end{align*}
Since the $\left\{  Y_{j}\right\}  $ are independent of each other and $\Delta
W,$ we have
\[
VarR_{1}=\sum_{k_{1},k_{2}=0}^{\infty}p\left(  k_{1},k_{2}\right)  \left(
\left(  \frac{\sigma_{01}}{2}\right)  ^{2}+k_{1}\sigma_{1}^{2}\right)  .
\]
Now using the fact that the mean and variance of $p_{k_{1}},$ defined in
$\left(  \ref{Pois}\right)  ,$ are both $\lambda_{1}\Delta t,$ one has
\[
\sum_{k_{2}=0}^{\infty}p\left(  k_{1},k_{2}\right)  =p_{k_{1}}=e^{-\lambda
_{1}\Delta t}\frac{\left(  \lambda_{1}\Delta t\right)  ^{k_{1}}}{k_{1}!},
\]%
\begin{equation}
VarR_{1}=\left(  \frac{\sigma_{01}}{2}\right)  ^{2}+\sigma_{1}^{2}\lambda
_{1}\Delta t, \label{V1}%
\end{equation}
and similarly for $VarR_{2}.$

Next, abbreviating $\mathbb{E}\left[  R_{1}R_{2}|\ \Delta N_{t}=k_{1}%
,\ \Delta\tilde{N}_{t}=k_{2}\right]  =\mathbb{E}\left[  R_{1}R_{2}%
|\ k_{1},k_{2}\right]  ,$ we calculate,
\begin{align*}
\mathbb{E}\left[  R_{1}R_{2}\right]   &  =\sum_{k_{1},k_{2}=0}^{\infty
}p\left(  k_{1},k_{2}\right)  \mathbb{E}\left[  R_{1}R_{2}|\ k_{1}%
,k_{2}\right] \\
&  =\sum_{k_{1},k_{2}=0}^{\infty}p\left(  k_{1},k_{2}\right)  \iint
\limits_{-\infty}^{\ \ \infty}x_{1}x_{2}f^{\left(  k_{1},k_{2}\right)
}\left(  x_{1},x_{2}\right)  dx_{1}dx_{2}%
\end{align*}
where $f^{\left(  k_{1},k_{2}\right)  }\left(  x_{1},x_{2}\right)  $ is the
joint density for $\hat{R}_{1},\hat{R}_{2}$ defined in $\left(  \ref{R1R2h}%
\right)  $ The covariance can then be expressed by writing%
\[
\mathbb{E}R_{1}\ \mathbb{E}R_{2}=\mu_{01}\mu_{02}=\sum_{k_{1},k_{2}=0}%
^{\infty}p\left(  k_{1},k_{2}\right)  \mu_{01}\mu_{02}%
\]%
\[
Cov\left(  R_{1},R_{2}\right)  =\sum_{k_{1},k_{2}=0}^{\infty}p\left(
k_{1},k_{2}\right)  \left(  \iint\limits_{-\infty}^{\ \ \infty}x_{1}%
x_{2}f^{\left(  k_{1},k_{2}\right)  }\left(  x_{1},x_{2}\right)  dx_{1}%
dx_{2}-\mu_{01}\mu_{02}\right)  .
\]
The term in the parentheses is $Cov\left(  \hat{R}_{1}\left(  k_{1}\right)
,\hat{R}_{2}\left(  k_{2}\right)  \right)  =\rho\left(  k_{1},k_{2}\right)
\sigma_{R_{1}}\left(  k_{1}\right)  \sigma_{R_{2}}\left(  k_{2}\right)  $ (see
$\left(  \ref{Cov}\right)  $ in Section 2) so we can write%
\[
Cov\left(  R_{1},R_{2}\right)  =\sum_{k_{1},k_{2}=0}^{\infty}p\left(
k_{1},k_{2}\right)  \rho\left(  k_{1},k_{2}\right)  \sigma_{R_{1}}\left(
k_{1}\right)  \sigma_{R_{2}}\left(  k_{2}\right)  .
\]
We can then write the overall correlation, $\rho_{T}$, as
\begin{align}
\rho_{T}  &  :=\frac{Cov\left(  R_{1},R_{2}\right)  }{\sqrt{VarR_{1}}%
\sqrt{VarR_{2}}}\label{st}\\
&  =\frac{\sum_{k_{1},k_{2}=0}^{\infty}p\left(  k_{1},k_{2}\right)
\rho\left(  k_{1},k_{2}\right)  \sigma_{R_{1}}\left(  k_{1}\right)
\sigma_{R_{2}}\left(  k_{2}\right)  }{\sqrt{\left(  \frac{\sigma_{01}}%
{2}\right)  ^{2}+\sigma_{1}^{2}\lambda_{1}\Delta t}\sqrt{\left(  \frac
{\sigma_{02}}{2}\right)  ^{2}+\sigma_{2}^{2}\lambda_{2}\Delta t}}.\nonumber
\end{align}
In order to show that $\rho_{T}=1$ can be unity when the correlation between
the bivariate Poisson processes, $\Delta N_{t}$ and $\Delta\tilde{N}_{t}$ is
one, we note first the result below, which follows immediately from the
definition, $\left(  \ref{bivar}\right)  .$ If the correlation between $\Delta
N_{t}$ and $\Delta\tilde{N}_{t}$ approaches $1$ then one must have
$\lambda_{01}\rightarrow0$, $\lambda_{02}\rightarrow0,$ while $\lambda
_{12}\rightarrow1,$ so that
\[
\lim_{\lambda_{0j}\rightarrow0}p\left(  k_{1},k_{2}\right)  =\left\{
\begin{array}
[c]{ccc}%
0 & if & k_{1}\not =k_{2}\\
e^{-\lambda_{12}}\frac{\lambda_{11}^{k_{1}}}{k_{1}!} & if & k_{1}=k_{2}%
\end{array}
\right.
\]
and thus, $p\left(  k_{1},k_{2}\right)  $ reduces to $p\left(  k_{1}%
,k_{2}\right)  =p_{k_{1}}\delta\left(  k_{1},k_{2}\right)  $ as the
correlation approaches $1.$

In the limit as the correlation between $\Delta N_{t}$ and $\Delta\tilde
{N}_{t}$ approaches $1,$ the numerator of $\left(  \ref{st}\right)  $ has the form:%

\begin{align*}
&  \sum_{k_{1},k_{2}=0}^{\infty}p\left(  k_{1},k_{2}\right)  \rho\left(
k_{1},k_{2}\right)  \sigma_{R_{1}}\left(  k_{1}\right)  \sigma_{R_{2}}\left(
k_{2}\right) \\
&  =\sum_{k_{1}=0}^{\infty}p_{k_{1}}\rho\left(  k_{1},k_{1}\right)
\sigma_{R_{1}}\left(  k_{1}\right)  \sigma_{R_{2}}\left(  k_{1}\right)  .
\end{align*}
If $\sigma_{01}=\sigma_{02}$ and $\sigma_{1}=\sigma_{2}$ (so the two types of
variances for the supply and demand are equal), and $\rho\left(  k_{1}%
,k_{2}\right)  =1$ then the numerator reduces to
\begin{align*}
\sum_{k_{1}=0}^{\infty}p_{k_{1}}\sigma_{R_{1}}\left(  k_{1}\right)
\sigma_{R_{2}}\left(  k_{1}\right)   &  =\sum_{k_{1}=0}^{\infty}p_{k_{1}%
}\left\{  \left(  \frac{\sigma_{01}}{2}\right)  ^{2}+\sigma_{1}^{2}k_{1}\Delta
t\right\} \\
&  =\left(  \frac{\sigma_{01}}{2}\right)  ^{2}+\sigma_{1}^{2}\lambda_{1}\Delta
t
\end{align*}
where we have have used the fact that the mean of $\Delta N_{t}$ is
$\lambda_{1}\Delta t,$ yielding the following result.

\bigskip\ \ 

\textbf{Theorem C.1.} Suppose that $\sigma_{01}=\sigma_{02}$ and $\sigma
_{1}=\sigma_{2}$, the correlation between $\Delta N_{t}$ and $\Delta\tilde
{N}_{t}$ approaches $1,$ and the covariance $\rho\left(  k_{1},k_{2}\right)  $
is set as an arbitrary number $r\in\left(  -1,1\right)  .$ Then $\rho
_{T}\rightarrow r.$

\bigskip

From $\left(  \ref{st}\right)  $ it is clear --as one would expect -- that one
cannot attain the full range $\left(  -1,1\right)  $ of $\rho_{T}$ if the
number of jumps in supply and demand are not correlated. Equation $\left(
\ref{st}\right)  $ can also be used to obtain the correlation when these two
Poisson processes are independent. Using the standard result that independence
implies%
\[
p\left(  k_{1},k_{2}\right)  =p_{k_{1}}p_{k_{2}}%
\]
one has the correlation for the overall probability as%

\begin{equation}
\rho_{T}=\frac{\sum_{k_{1},k_{2}=0}^{\infty}p_{k_{1}}p_{k_{2}}\rho\left(
k_{1},k_{2}\right)  \sigma_{R_{1}}\left(  k_{1}\right)  \sigma_{R_{2}}\left(
k_{2}\right)  }{\sqrt{\left(  \frac{\sigma_{01}}{2}\right)  ^{2}+\sigma
_{1}^{2}\lambda_{1}\Delta t}\sqrt{\left(  \frac{\sigma_{02}}{2}\right)
^{2}+\sigma_{2}^{2}\lambda_{2}\Delta t}} \label{ind}%
\end{equation}
Since $\sigma_{R_{1}}^{2}\left(  k_{1}\right)  =\left(  \frac{\sigma_{01}}%
{2}\right)  ^{2}+k_{1}\sigma_{1}^{2}$ and likewise for $\sigma_{R_{2}}%
^{2}\left(  k_{2}\right)  $, one has the following exact total correlation
under the condition that the Poisson processes are independent but the
distribution of the jumps have covariance $\sigma\left(  k_{1},k_{2}\right)
:$
\[
\rho_{T}=\frac{\sum_{k_{1},k_{2}=0}^{\infty}p_{k_{1}}p_{k_{2}}\sigma\left(
k_{1},k_{2}\right)  \sqrt{\left(  \frac{\sigma_{01}}{2}\right)  ^{2}%
+k_{1}\sigma_{1}^{2}}\sqrt{\left(  \frac{\sigma_{02}}{2}\right)  ^{2}%
+k_{2}\sigma_{2}^{2}}}{\sqrt{\left(  \frac{\sigma_{01}}{2}\right)  ^{2}%
+\sigma_{1}^{2}\lambda_{1}\Delta t}\sqrt{\left(  \frac{\sigma_{02}}{2}\right)
^{2}+\sigma_{2}^{2}\lambda_{2}\Delta t}}.
\]
Note that even if we set $\sigma_{01}=\sigma_{02}$ and $\sigma_{1}=\sigma
_{2},$ and $\rho\left(  k_{1},k_{2}\right)  =1$ we obtain%
\[
\rho_{T}=\frac{\left(
{\displaystyle\sum\limits_{k_{1}=0}^{\infty}}
p_{k_{1}}\sqrt{\left(  \frac{\sigma_{01}}{2}\right)  ^{2}+k_{1}\sigma_{1}^{2}%
}\right)  ^{2}}{\left(  \frac{\sigma_{01}}{2}\right)  ^{2}+\sigma_{1}%
^{2}\lambda_{1}\Delta t}.
\]
The denominator can be written as $\sum_{k_{1}=0}^{\infty}p_{k_{1}}\left\{
\left(  \frac{\sigma_{01}}{2}\right)  ^{2}+\sigma_{1}^{2}k_{1}\right\}  ,$ so
that the Cauchy-Schwarz inequality implies $\rho_{T}\leq1$ with equality
holding only for trivial Poisson probabilities $\left\{  p_{k_{1}}\right\}  .$

This shows that when the Poisson processes are uncorrelated the minimum of
$\rho_{T}$ even with full anti-correlation, $\sigma\left(  k_{1},k_{2}\right)
=-1$ will exceed $-1.$ However, when there is a bivariate Poisson process, the
value of $\rho_{T}$ can span the range $\left(  -1,1\right)  $ depending on
the $\sigma\left(  k_{1},k_{2}\right)  .$

\bigskip

\textbf{Appendix D. Randomness and the Central Limit Theorem in supply and
demand.} We consider a large number of agents buying an asset, and argue that
the random orders satisfy the the Central Limit Theorem (CLT), and model the
price change within a discrete formulation of the basic equation%

\[
P^{-1}\frac{dP}{dt}=\frac{\mathcal{D}\left(  t,P\left(  t,\Gamma\right)
,\omega\right)  }{\mathcal{S}\left(  t,P\left(  t,\bar{\Gamma}\right)
,\bar{\omega}\right)  }-1.
\]
We consider a discrete set of times, $\left\{  t_{j}\right\}  _{j=1}^{\infty
},$ and prices, $\left\{  P_{i}\right\}  _{i=1}^{N}.$ We let $\Gamma
_{j}:=\left\{  \omega_{0},...,\omega_{j},\bar{\omega}_{0},...,\bar{\omega}%
_{j}\right\}  $\ where $\omega\in\Omega,$ $\bar{\omega}\in\Omega$ \ are random
values that will be chosen from a distribution. When the computed price is not
exactly one of those values, we approximate with the closest of the $P_{i}.$
The demand at the time $t_{j}$ can depend on the existing price, $P\left(
t_{j},\Gamma_{j}\right)  $ and on $t_{j}$ directly, due to a multitude of
factors. In addition, the demand will depend on a random value that depends on
$\omega_{j}.$ The situation is the same for supply. The randomness can depend
on the time and price in a complicated way. As noted earlier, script
$\mathcal{D},$ is used to indicate the full demand (including randomness)
while $D$ is used for the deterministic part, or equivalently the expected
value of demand (assuming the random term has vanishing mean). We will thus
assume $\mathbb{E}\mathcal{D=}D,$ and analogously for supply.

We illustrate the idea with a simple model below.

\textbf{A basic model.} One simple possibility is that the random variable,
say, $R_{j}\left(  \omega_{j}\right)  $, alters the deterministic demand,
$D\left(  t,P\left(  t,\Gamma\right)  \right)  $ via%
\begin{equation}
\mathcal{D}\left(  t_{j},P\left(  t_{j},\Gamma_{j}\right)  ,\omega_{j}\right)
=D\left(  t_{j-1},P\left(  t_{j-1},\Gamma_{j-1}\right)  \right)  \left(
1+R_{j}\left(  \omega_{j}\right)  \right)  , \label{dr}%
\end{equation}
and analogously for supply with $\bar{R}_{j}\left(  \bar{\omega}_{j}\right)
.$ In other words, at time $t_{j}$ we have a particular price, $P\left(
t_{j-1},\Gamma_{j-1}\right)  $ that depends on the random variables chosen up
through time $t_{j-1},$ i.e., $\Gamma_{j-1}.$ If there were no additional
randomness at time $t_{j}$ we would have the total demand, $\mathcal{D},$ that
would lack the $R_{j}\left(  \omega_{j}\right)  $ term. Randomness introduced
in this multiplicative form simply states that the magnitude of the randomness
is proportional to the overall demand. For example, if the expectation of
demand is a million units, we would expect that the standard deviation would
be much larger than if it were a hundred units. Thus, simply adding $R_{j}$ to
the demand would not be meaningful.

Next, we consider the distribution of $R_{j}.$ For an actively traded stock,
we can assume that there are many independent traders and investors that can
be approximated by some distribution. If the number of such traders is
sufficiently large, the distribution of the sum can be approximated well by
the normal distribution. This is a consequence of the Central Limit Theorem
(CLT) which is applied here directly to the large number of agents
(individuals or institutions). Consider, for example, $1000$ agents bidding on
an asset at a particular time with a particular distribution that is not
necessarily normal. For a broad set of distributions, the CLT will apply, and
we can thus assume that the aggregate demand is normal. This is in sharp
contrast to trying to apply CLT directly to prices, since prices are not
chosed directly by agents, but arise from a complex process. As shown in the
literature \cite{DR}, \cite{GE} even the quotient of two normals is not
normal, though in some regimes of parameters, e.g., near the mean, the
distribution of the quotient can be approximated by a Gaussian distribution.

The price will depend on the set of $\omega_{i},$ $\bar{\omega}_{i}$ prior to
that time, i.e., $\Gamma_{i}$.$\ \ $Let%
\[
\Delta P\left(  t_{j},\Gamma_{j}\right)  :=P\left(  t_{j},\Gamma_{j}\right)
-P\left(  t_{j-1},\Gamma_{j-1}\right)  .
\]

The discrete version of the price equation can be written, with $R_{j+1}$
normal,%
\[
\frac{\Delta P\left(  t_{j+1},\Gamma_{j+1}\right)  }{P\left(  t_{j},\Gamma
_{j}\right)  }=\frac{D\left(  t_{j},P\left(  t_{j},\Gamma_{j}\right)  \right)
\left(  1+R_{j+1}\left(  \omega_{j+1}\right)  \right)  }{S\left(
t_{j},P\left(  t_{j},\Gamma_{j}\right)  \right)  \left(  1+\bar{R}%
_{j+1}\left(  \bar{\omega}_{j+1}\right)  \right)  }-1
\]
so that as we obtain the successive terms,%
\[
P\left(  t_{j+1},\Gamma_{j+1}\right)  =\frac{D\left(  t_{j},P\left(
t_{j},\Gamma_{j}\right)  \right)  \left(  1+R_{j+1}\left(  \omega
_{j+1}\right)  \right)  }{S\left(  t_{j},P\left(  t_{j},\Gamma_{j}\right)
\right)  \left(  1+\bar{R}_{j+1}\left(  \bar{\omega}_{j+1}\right)  \right)
}P\left(  t_{j},\Gamma_{j}\right)  .
\]

We note that in this simple model, the randomness influences the demand at all
prices uniformly through $\left(  \ref{dr}\right)  .$

\textbf{A more general model.} We now consider distributions that can depend
not only on time, but on the price as well. For example, one might have
greater variance for prices that are extreme. Changing notation, we focus on a
single time $t_{j}$ and write the random variables $\mathcal{D},$
$\mathcal{S}$ and $X$ as%
\begin{align*}
\mathcal{D}  &  =\left(  \mathcal{D}_{1},...,\mathcal{D}_{N}\right)
,\ \ \mathcal{S}=\left(  \mathcal{S}_{1},..,\mathcal{S}_{N}\right)  ,\\
X  &  =\left(  X_{1},..,X_{2N}\right)  =\left(  \mathcal{D}_{1}%
,...,\mathcal{D}_{N},\mathcal{S}_{1},..,\mathcal{S}_{N}\right)  ,
\end{align*}
where $\mathcal{D}_{k}$ is the demand at price $P_{k}$. At this time $t_{j}$
there is a deterministic component (i.e., expected value) of the demand at
each price $P_{k}$ with $k=1,2,..,N$, denoted $\mu_{k}$ with $k=1,2,..,N.$
Similarly, for the supply, $X_{N+1},...X_{2N}$ one has $\mu_{N+1},..,\mu
_{2N}.$ The covariance matrix, $\Sigma,$ is arbitrary, provided it is positive
definite. Thus, one can specify the correlations between the supply and demand
at different prices. Note that the correlation between the supply and demand
at a given price will generally be negative, and close to $-1.$ Hence, the
density of $X$ is a multivariate normal described by%
\begin{align*}
f\left(  x;\mu,\Sigma\right)   &  =\left(  2\pi\right)  ^{-N}\left\vert
\Sigma\right\vert ^{-1/2}e^{-Q_{2N}\left(  x;\mu,\Sigma\right)  /2}\\
Q_{2N}\left(  x;\mu,\Sigma\right)   &  :=\left(  x-\mu\right)  ^{T}\Sigma
^{-1}\left(  x-\mu\right)  .
\end{align*}

The price change at time $t_{j+1}$ when the price $P\left(  t_{j},\Gamma
_{j}\right)  =P_{k}$ is thus given by%
\begin{align*}
\frac{\Delta P\left(  t_{j+1},\Gamma_{j+1}\right)  }{P\left(  t_{j},\Gamma
_{j}\right)  }  &  =\frac{\mathcal{D}_{k}\left(  t_{j}\right)  }%
{\mathcal{S}_{k}\left(  t_{j}\right)  }-1=\frac{X_{k}\left(  t_{j}\right)
}{X_{N+k}\left(  t_{j}\right)  }-1,\text{ \ i.e.,}\\
P\left(  t_{j+1},\Gamma_{j+1}\right)   &  =P\left(  t_{j},\Gamma_{j}\right)
\frac{X_{k}\left(  t_{j}\right)  }{X_{N+k}\left(  t_{j}\right)  }.
\end{align*}
If we consider the full nonlinear model with general $G$ subject to the
conditions stated earlier (including $G\left(  1\right)  =0,$ $G^{\prime}>0$)
the equation has the form
\[
P^{-1}\frac{dP}{dt}=G\left(  \frac{\mathcal{D}\left(  t,P\left(
t,\Gamma\right)  ,\omega\right)  }{\mathcal{S}\left(  t,P\left(  t,\bar
{\Gamma}\right)  ,\bar{\omega}\right)  }\right)  .
\]
The discrete model above can then be written as%
\[
P\left(  t_{j+1},\Gamma_{j+1}\right)  =\left[  G\left(  \frac{X_{k}\left(
t_{j}\right)  }{X_{N+k}\left(  t_{j}\right)  }\right)  +1\right]  P\left(
t_{j},\Gamma_{j}\right)  .
\]

\bigskip

\textbf{Appendix E. Issue of negative supply or demand. }We consider, for
simplicity, the situation without the jump terms as the issues are similar.

\bigskip

(I) Does negative demand make any sense in finance? While a standard
interpretation is that all buy orders are demand and all sell orders are
supply, there is also the concept of selling short, which could be interpreted
as negative demand, for example. Whether this is useful depends on the other
equations coupled with the price equation, as in the asset flow equations. In
other words, if there is a group of investors that is largely on the buy-side,
it may be a useful mathematical concept to consider selling short as negative
demand. Whether this is useful depends on the other equations coupled with the
price equation, as in the asset flow equations.

For commodities, negative price means that one has to pay someone to take it
away, as happened with oil prices in May 2020. This could be interpreted as
negative net demand.

In the text, we consider the mathematics in full generality as the
mathematical issue involving a quotient of such random variables in other contexts.

\bigskip

(II) The normal distribution is used in many measurable quantities that are
positive by definition, e.g. IQ, prices, etc. While negative values are
theoretically possible from the normal distribution, the standard deviation is
sufficiently small compared to the mean that the probability of a negative
value is infinitesimal. The standard equation for asset price, $\left(
\ref{cs}\right)  ,$ can lead to negative prices, for example. Indeed, letting
$\mu=0,$(i.e., no trend) and $\sigma=const$ for simplicity, and fixing a time
interval $\Delta t,$ so $\Delta W\sim\mathcal{N}\left(  0,\Delta t\right)  ,$
one has%
\[
\frac{\Delta P}{P}\tilde{=}\sigma\Delta W.
\]

Thus, one has%
\[
P\left(  t+\Delta t\right)  \tilde{=}P\left(  t\right)  \left(  1+\sigma\Delta
W\right)  .
\]
The probability of a negative price is the probability that a random variable
$X\sim\mathcal{N}\left(  0,\sigma^{2}\Delta t\right)  $ is less than $-1,$
i.e., this is a $\left(  \sigma\left(  \Delta t\right)  ^{1/2}\right)  ^{-1}$
standard deviation event. For the S\&P 500, or a typical stock, the standard
deviation of the price is about $0.5\%$ or $1\%$ of the price. \ Thus, we set
$\Delta t=1$ and $\sigma=0.01,$ so the equation above indicates (using $1\%$)
that the probability of going to negative price in one day is a $100$ standard
deviation event. Hence, this possibility is negligible within this framework.

The situation is similar for negative demand or supply, as one could verify --
in principle -- from the order book data. Also, this can be related to
relative price change using the price change equation $\left[  \ref{dpg}%
\right]  $, and shown to be the same order of magnitude as the probability of
negative prices.

\bigskip

(III) There is also the possibility of using conditional probability. This, of
course, changes the distribution by an infinitesimal amount (provided the
standard deviation is small compared to the mean), and it forces strictly
positive values of supply and demand. We can think of this as a random choice
of demand that is made from a normal distribution with mean $\mu_{\mathcal{D}%
}$ and variance $\sigma_{\mathcal{D}}^{2}.$ If the value is positive, it is
used. In the extremely rare event that it is negative, the choice is
discarded. When $\mu_{\mathcal{D}}/\sigma_{\mathcal{D}}$ exceeds $5$ or $6,$
the two densities would be nearly identical for practical purposes. However,
the analysis of this conditional probability is useful in other contexts
beyond price, demand and supply, particularly, when the standard deviation is
not small compared to the mean.

\bigskip

(IV) There are other ways to modify the normal distribution so that the random
variable has no negative values, e.g., by truncation. The resulting
distribution is of course not exactly normal, and in practice, the difference
between this approach and the conditional approach we have used\ would be
negligible in this application, but could be significant when the standard
deviation is not small compared to the mean.

\bigskip

\textbf{Acknowledgement.} The author thanks the Hayek Fund for Scholars for
its support.

\bigskip

\begin{center}
\bigskip\textbf{CAPTION FOR FIGURE 1}

\textbf{\bigskip}
\end{center}

The graph of $y=r\left(  x\right)  :=x-x^{-1}$ featuring the two branches is
displayed. Calculations of the density for $r\left(  R_{a}\right)  $ required
determining the intersection of a constant value of $y$ with $r\left(
x\right)  .$ When $y>0$ the intersection in the upper half plane occurs at the
values $x_{+}^{>}$ for the right branch, and $x_{-}^{>}$ for the left branch
on the $x-$axis. Analogously the intersections for the lower half plane
($y<0$) occur at $x_{+}^{<}$ and $x_{-}^{<}.$ The qualitative features of
$G_{\varepsilon}\left(  x\right)  $ are similar, and the same notation is used
when $G_{\varepsilon}\left(  x\right)  $ replaces $r\left(  x\right)  .$

\end{document}